\documentclass[12pt]{article}
\usepackage{latexsym,amsmath,amssymb,amsbsy,graphicx}
\usepackage[cp1251]{inputenc}
\usepackage[T2A]{fontenc}
\usepackage[russian,english]{babel}
\usepackage[space]{cite} 

\begin{document}

\bigskip
\centerline {\bf Evolution of the cycles of magnetic activity of the
Sun} \centerline {\bf and Sun-like stars in time}
\bigskip

\centerline {E.A. Bruevich $^{a}$ , V.V. Bruevich $^{b}$, B.P.
Artamonov $^{c}$ }

\centerline {\it $^{a,b,c}$ Sternberg Astronomical Institute, Moscow
State
 University,}
\centerline {\it Universitetsky pr., 13, Moscow 119992, Russia}\

\centerline {\it e-mail:  $^a${red-field@yandex.ru},
$^b${brouev@sai.msu.ru}, $^c${artamon@sai.msu.ru}}\ {\bf Abstract}

We applied the method of continuous wavelet-transform to the
time-frequency analysis to the sets of observations of relative
sunspot numbers, sunspot areas and to 6 Mount Wilson HK-project
stars with well-defined magnetic cycles. Wavelet analysis of these
data reveals the following pattern: at the same time there are
several activity cycles whose periods vary widely from the
quasi-biennial up to the centennial period for the Sun and vary
significant during observations time of the HK-project stars. These
relatively low-frequency periodic variations of the solar and
stellar activity gradually change the values of periods of different
cycles in time. This phenomenon can be observed in every cycles of
activity.

\bigskip

\vskip12pt
\section{Introduction}
\vskip12pt

The study of magnetic activity of the Sun and the stars which is
called by the complex of electromagnetic and hydrodynamic processes
in their atmospheres is of fundamental importance for astrophysics.

The relative sunspot number SSN is a very popular, widely used solar
activity index: the series of relative sunspot numbers direct
observations continue almost two hundred years. The SSN index has an
advantage over other indices of activity because data on annual
variation available from the 1700's: for 1700--1849 on the basis of
undirect data, later then 1850 -- according to the direct
observations. The SSN (also known as the International sunspot
number, relative sunspot number, or Wolf number) is a quantity that
measures the number of sunspots and groups of sunspots present on
the surface of the sun.
 Sunspots are temporary
phenomena on the photosphere of the Sun that appear visibly as dark
spots compared to surrounding regions. They are caused by intense
magnetic activity, which inhibits convection by an effect comparable
to the eddy current brake, forming areas of reduced surface
temperature. Although they are at temperatures of roughly
3000-4500K), the contrast with the surrounding material at about
5.780 K leaves them clearly visible as dark spots. Manifesting
intense magnetic activity, sunspots host secondary phenomena such as
coronal loops(prominences) and reconnection events. Most of solar
flares and coronal mass ejections originate in magnetically active
regions around visible sunspot groupings. Similar phenomena
indirectly observed on stars are commonly called stars pots and both
light and dark spots have been measured. Thus the cyclic variations
of the SSN and the evolution of these cycles in time is an important
task for the study of all complex phenomena on the Sun associated
with different indices of solar activity.

The historical sunspot record was first put by Wolf in 1850s and has
been continued later in the 20th century until today. Wolf's
original definition of the relative sunspot number for a given day
as $R = 10 \cdot$ Number of Groups + Number of Spots  visible on the
solar disk has stood the test of time. The factor of 10 has also
turned out to be a good choice as historically a group contained on
average ten spots. Almost all solar indices and solar wind
quantities show a close relationship with the SSN, see [1],[2].

We have to point out that close interconnection between radiation
fluxes characterized the energy release from different atmosphere's
layers is the widespread phenomenon among the stars of late-type
spectral classes, see [3]. It was confirmed that there exists the
close interconnection between photospheric and coronal fluxes
variations for Sun-like stars of F, G, K and M spectral classes with
widely varying activity of their atmospheres, see [4],[5]. It was
also shown that the summary areas of spots and values of X-ray
fluxes increase gradually from the sun and Sun-like Mount Wilson HK
project stars [6] with the low spotted discs to the highly spotted K
and M-stars. The main characteristic describing the photospheric
radiation is the spottiness of the stars. Thus, the study of the
relative sunspot numbers is very important to explain the
observations of sun-like stars.

The level of chromospheric activity of the Sun is consistent with
that of HK-project stars, which have well-defined cycles of
activity, but the level of coronal activity of the Sun
 is significantly below that of the coronal activity of Sun-like G-stars
from the different observational Programs which are studied Sun-like
stars: (1) HK-project -- the Mount Wilson program, see [6]; (2) The
California \& Carnegie Planet Search Program which includes
observations of approximately 1000 stars at Keck \& Lick
observatories  in chromospheric CaII H\&K emission cores, see [7];
(3) The Magellan Planet Search Program which includes Las Campanas
Observatory CA measurements of 670 F, G, K and M main sequence stars
of the Southern Hemisphere. $S_{HK}$-indexes of these stars are also
converted to the Mount Wilson system, see [8].

A comparative analysis of chromospheric, coronal and cyclic activity
of the Sun and Sun-like stars of F, G and K spectral classes from
these different observational Programs shows the similar
characteristics of magnetic cycles on the Sun and on the Sun-like
stars, see [9],[10].

We have studied: (1) -- yearly averaged values of SSN during solar
activity cycles 1 -- 23, the tree-hundred yrs data set; (2) --
yearly averaged values of sunspot areas A, the 400 yrs data set, see
[11]; (3) -- monthly averaged values of SSN during activity cycles
18 -- 24 and (4)-- daily averaged values of SSN during activity
cycle 22. All SSN data are available at NGDC web site, see
observational data from National Geophysical Data Center. Solar Data
Service [12].

For the HK-project stars study we have applied the wavelet analysis
for partially available data from the records of relative CaII
emission fluxes - the variation of $S_{HK}$-indexes for 1965--1992
observation sets from Baliunas et al. (1995) and for 1985-- 2002
observations from [13]. We used the detailed plots of
$S_{HK}$-indexes time dependencies: each point of the record of
observations, which we processed in this paper using wavelet
analysis technique, corresponds to three months averaged values of
$S_{HK}$.

\vskip12pt
\section{Wavelet-analysis of series of observations of SSN}
\vskip12pt

\begin{figure}[h!!!]
   \centerline
   {\includegraphics[width=8.0cm, angle=0]{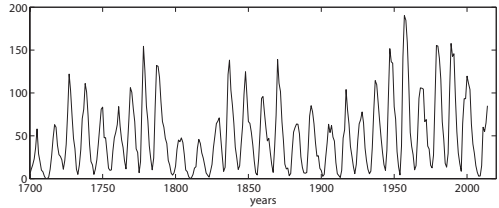}}
\caption{The time series of the annual averaged values of sunspot
numbers SSN from 1700 to 2014.}
  \label{Fig1}
   \end{figure}

\begin{figure}[h!!!]
   \centerline
   {\includegraphics[width=8.0cm, angle=0]{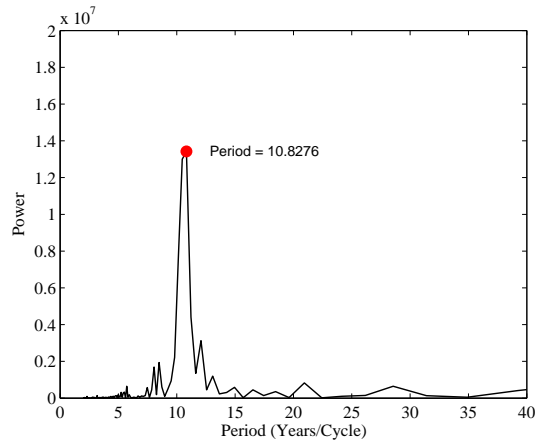}}
\caption{The Fast Fourier Transform of the time series of the annual
averaged values of sunspot numbers SSN from 1700 to 2014. The red
dot located the precise pick of the strongest frequency.}
  \label{Fig2}
   \end{figure}

\begin{figure}[h!]
 \centerline{\includegraphics[width=140mm]{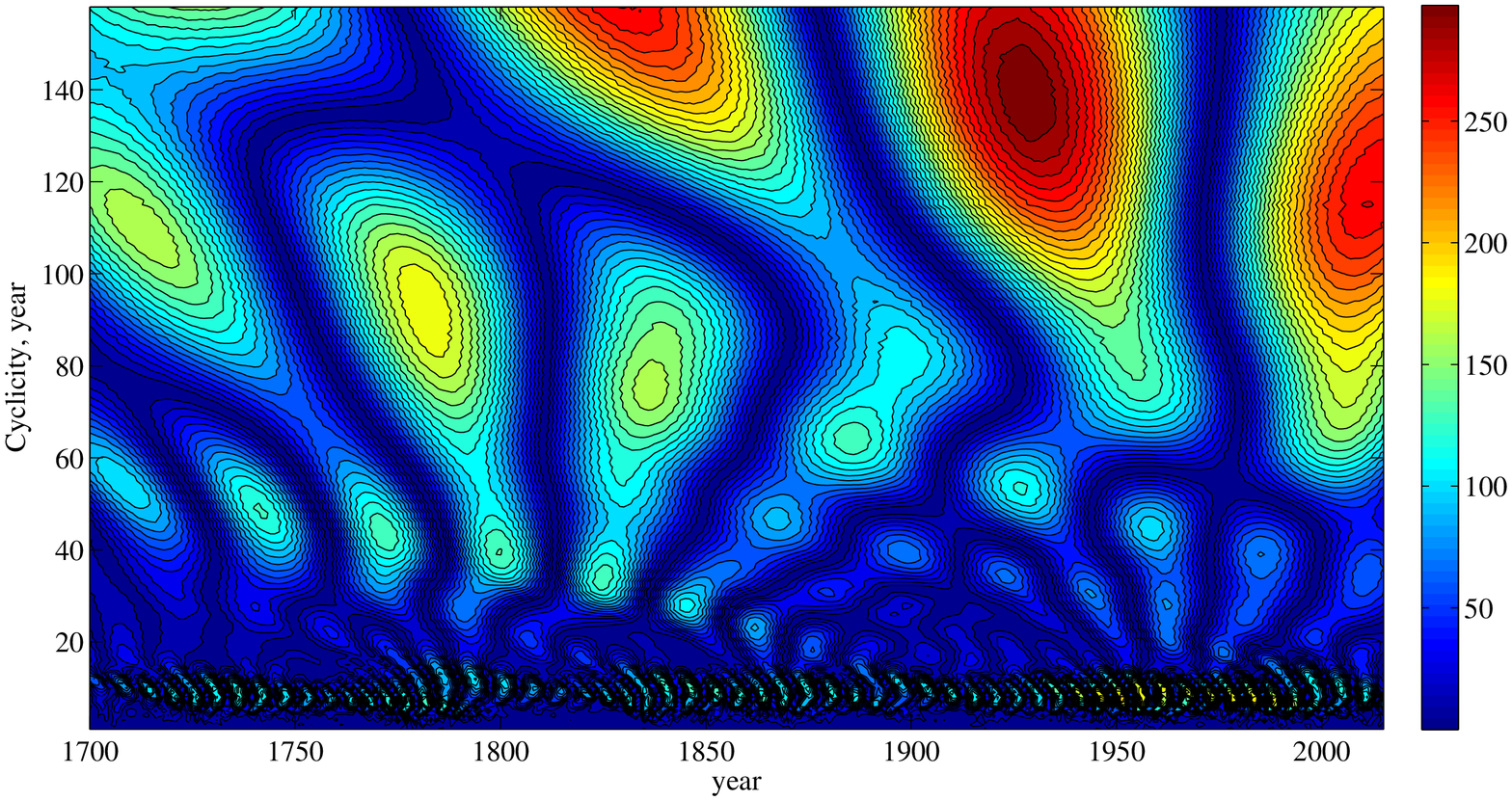}}
\caption{The wavelet image of the time series of the annual averaged
values of sunspot numbers SSN from 1700 to 2014 (for wavelet
transform calculations we used the Daubechies 10 mother wavelet).}
\label{Fig3}
\end{figure}

In Figure 1 we can see that the duration of the 11 yr cycle of solar
activity ranged from 7 to 17 yrs. The results of observation become
more accurate with the beginning of the of direct solar observations
(1850--2015).

In Figure 2 we can see how to use Fast Fourier Transform (FFT)
method to analyze the variations in sunspot activity (SSN) over the
last 300 yr. We can see that the FFT method gives only the value of
the duration of the main cycle of magnetic activity of the Sun,
which according to the 300-yr observations is equal to 10.83 years.
This method does not show cycle's period evolution: it's known that
according to the last 100-year observations in the XX century, the
value of the duration of the main cycle of magnetic activity of the
Sun is equal to 10.2 yrs, see [14].

In Figure 3 we illustrated with help of wavelet - analysis the fact
that the long time series of observations give us the very useful
information for study of the problem of solar flux cyclicity on long
time scales. The result of wavelet - analysis (Daubechies wavelet)
of series of observations of average annual SSN
 is presented in form of many of isolines. For the
isoline of the value of the wavelet-coefficients are of the same.
The maximum values of isolines specify the maximum values of
wavelet-coefficients, which corresponds to the most likely value of
the period of the cycle. We see there three well-defined cycles of
activity: - the main cycle of activity (this cycle is approximately
equal to a 10 - 11 yrs), 40-50- yr cyclicity and 100 -- 120 yr
(ancient) cyclicity.

In Figure 3 we can see that periods of cycles on different time
scales are not constant.

\begin{figure}[h!!!]
   \centerline
   {\includegraphics[width=8.0cm, angle=0]{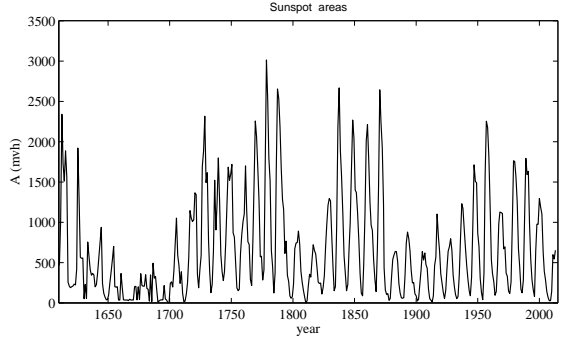}}
\caption{The time series of the annual averaged values of the areas
of sunspot numbers A (mvh) from 1610 to 2014.}
  \label{Fig4}
   \end{figure}

\begin{figure}[h!!!]
   \centerline
   {\includegraphics[width=8.0cm, angle=0]{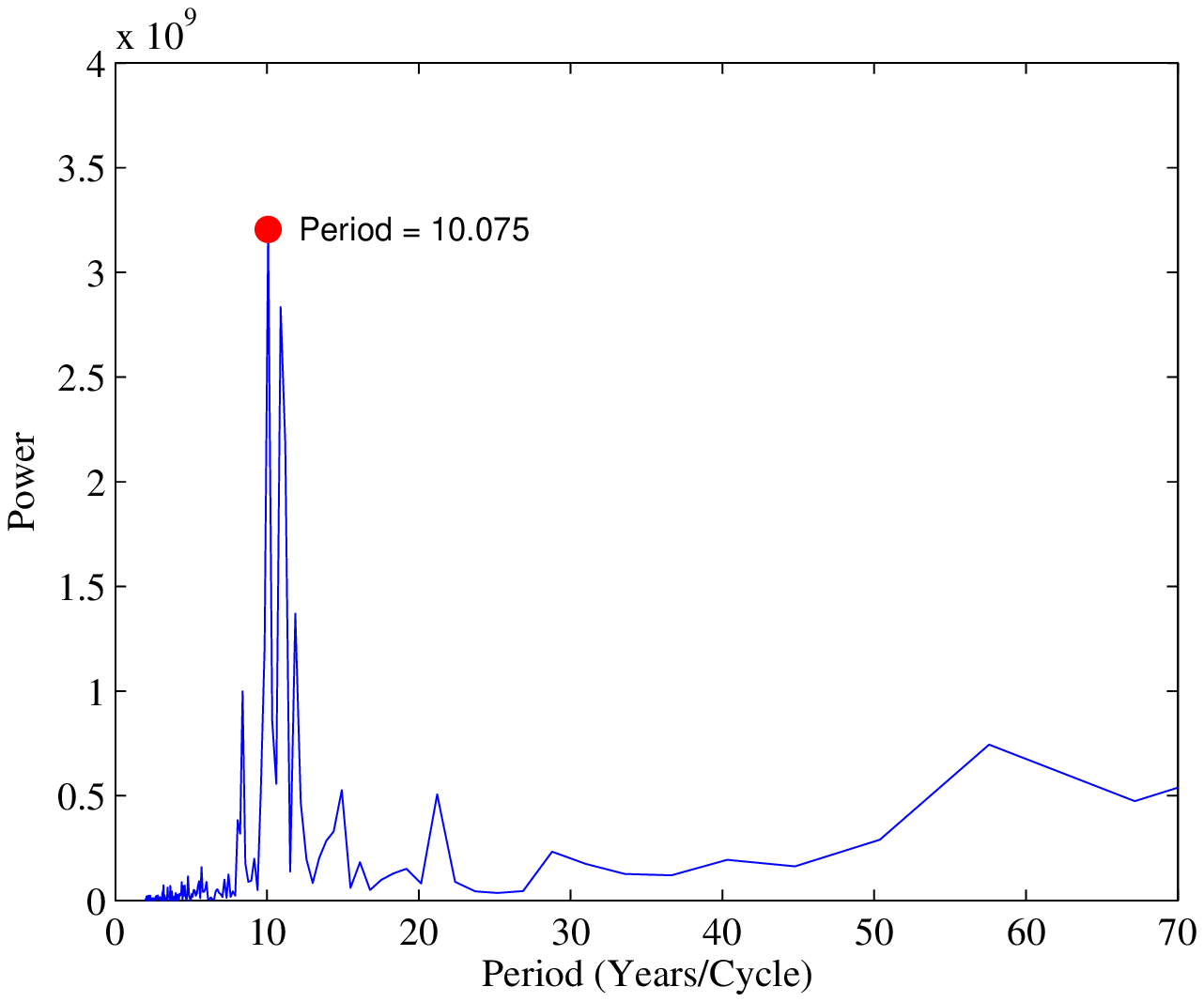}}
\caption{The Fast Fourier Transform of the time series of the annual
averaged values of the areas of sunspot numbers A (mvh) from 1610 to
2014. The red dot located the precise pick of the strongest
frequency.}
  \label{Fig5}

 \centerline{\includegraphics[width=140mm]{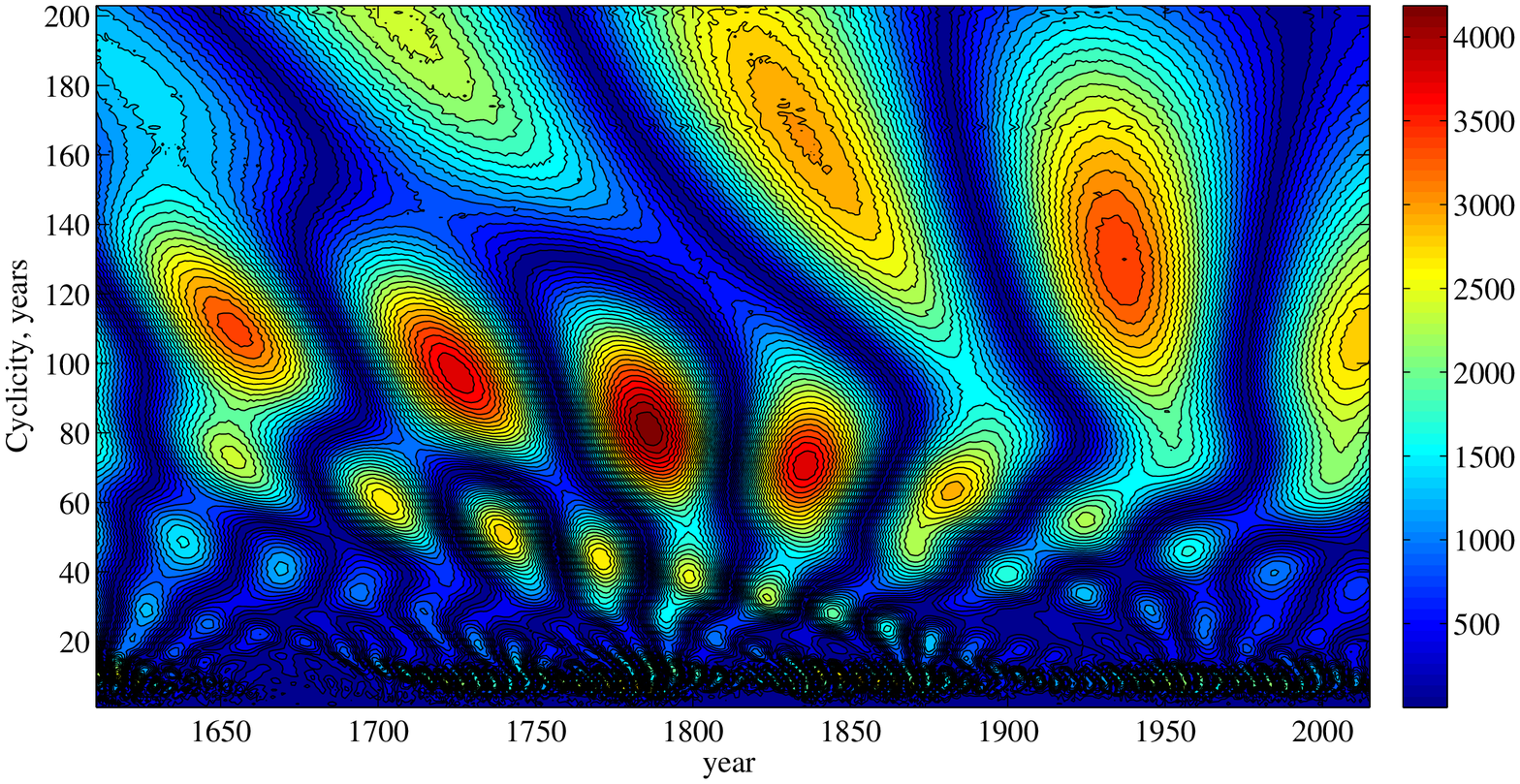}}
\caption{The wavelet image of the time series of the annual averaged
values of the areas of sunspot numbers A (mvh) from 1610 to 2014
(for wavelet transform calculations we used the Daubechies mother
wavelet).}
\label{Fig6}
\end{figure}

In Figure 4 we can see that the duration of the 11 yr cycle of solar
activity according to the annual averaged values of the areas of
sunspot numbers A (mvh) from 1610 to 2014 are also ranged from 7 to
17 yrs.

Figure 5 shows the Fast Fourier Transform (FFT) method of analyzing
the variations in solar activity according to the annual averaged
values of the areas of sunspot numbers A (mvh) over the last 400
year. The FFT gives only the value of the duration of the main cycle
of magnetic activity of the Sun, which according to the 400-year
observations is equal to 10.075 yrs. This method also does not show
cycle's period evolution.

In Figure 6 we also illustrated with help of wavelet - analysis the
fact that the long time series of observations give us the very
useful information for study of the problem of solar flux cyclicity
on long time scales. The result of wavelet - analysis (Daubechies
wavelet) of series of observations of average annual A (mvh)
 is presented in form of many of isolines. We also can see in Figure 6 three well-defined cycles of
activity: - the main cycle of activity (this cycle is approximately
equal to a 10 - 11 yrs), 40-50-yr cyclicity and 100 to 120-yr
(ancient) cyclicity.

For the wavelet analysis of relative sunspot numbers on the scales
in 11 years and quasi-biennial scales we will use the monthly
averaged values of SSN, see Figure 7.

\begin{figure}[tbh!]
\centerline{
\includegraphics[width=80mm]{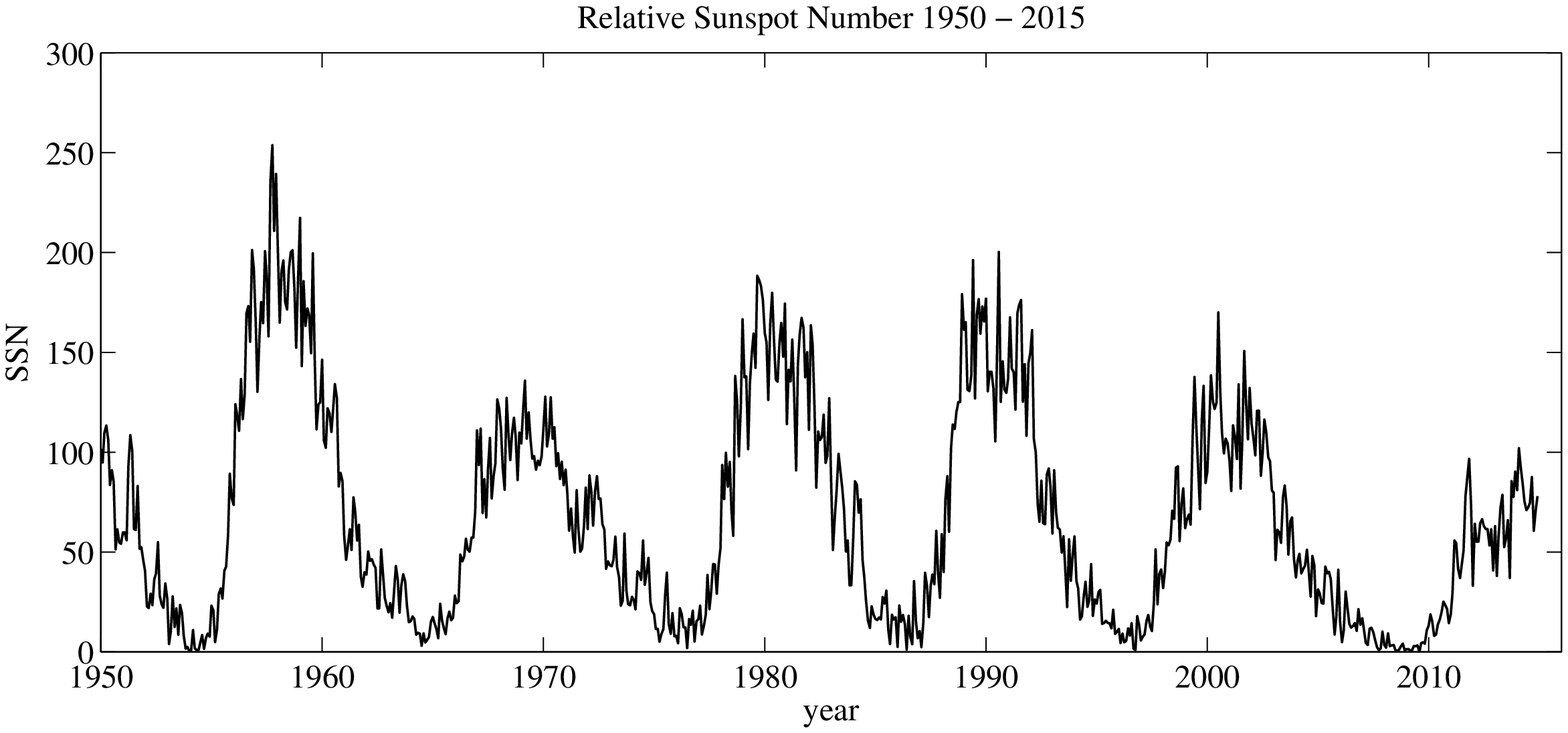}}
 \caption{Monthly averaged relative sunspot numbers SSN 1950 - 2015.}
\label{Fig7}
\end{figure}

\begin{figure}[tbh!]
\centerline{
\includegraphics[width=120mm]{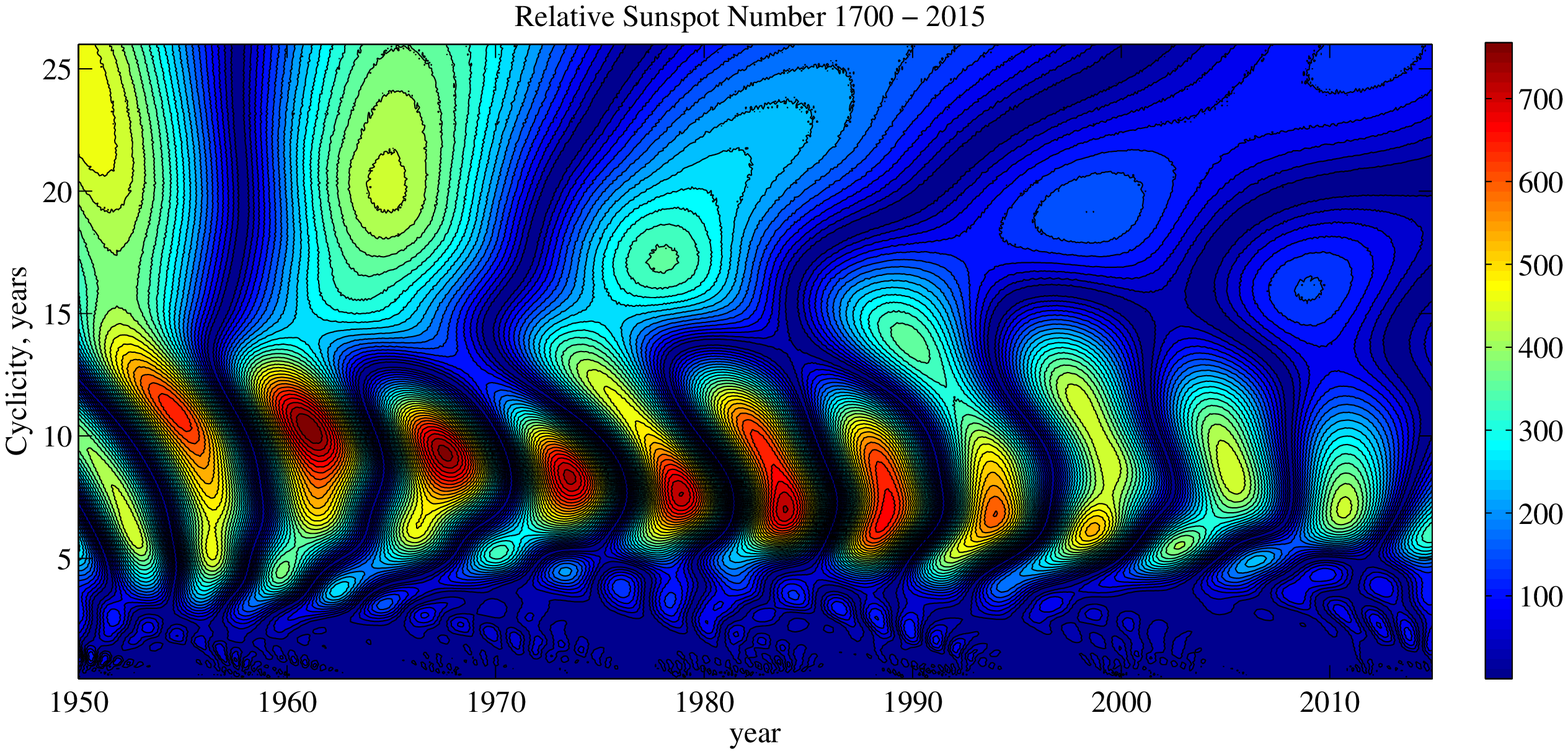}}
 \caption{The wavelet image of the cyclic activity of relative sunspot numbers SSN in  activity cycles 18 - 24.}
\label{Fig8}
\end{figure}

The long-term behavior of the sunspot group numbers have been
analyzed using wavelet technique by [15] who plotted the changes of
the Schwabe cycle (its period is about 11-yr) and studied the grand
minima. The temporal evolution of the Gleissberg cycle (its period
is about 100-yr) can also be seen in the time-frequency distribution
of the solar data. According to [15] the Gleissberg cycle is as
variable as the Schwabe cycle. It has two higher amplitude
occurrences: first around 1800 (during the Dalton minimum), and then
around 1950. They found very interesting fact - the continuous
decrease in the frequency (increase of period) of Gleissberg cycle.
While near 1750 the cycle length was about 50 yr, it lengthened to
approximately 130 yr by 1950.

The study of the indices of solar activity are very important not
only for analysis of solar radiation which comes from different
altitudes of solar atmosphere. The most important for
solar-terrestrial physics is the study of solar radiation influence
on the different layers of terrestrial atmosphere (mainly the solar
radiation in EUV/UV spectral range which effectively heats the
thermosphere of the Earth and so affects to our climate).

In the late of XX century some of solar physicists began to examine
with different methods the variations of relative sunspot numbers
not only in high amplitude 11-yr Schwabe cycle but in low amplitude
cycles approximately equal to half (5.5-yr) and fourth
 (quasi-biennial) parts of period of the main 11-yr cycle, see [16].
The periods of the quasi-biennial cycles vary considerably within
one 11-yr cycle, decreasing from 3.5 to 2 yrs, and this fact
complicates the study of such periodicity using the method of
periodogram estimates.

Using the methods of frequency analysis of signals  the
quasi-biennial cycles have been  studied not only for the relative
sunspot number, but also for 10.7 cm solar radio emission and for
some other indices of solar activity, see [17].

 It was also shown that the cyclicity
on the quasi-biennial time scale takes place often among the stars
with 11-yr cyclicity, see [14].

\begin{figure}[tbh!]
\centerline{
\includegraphics[width=80mm]{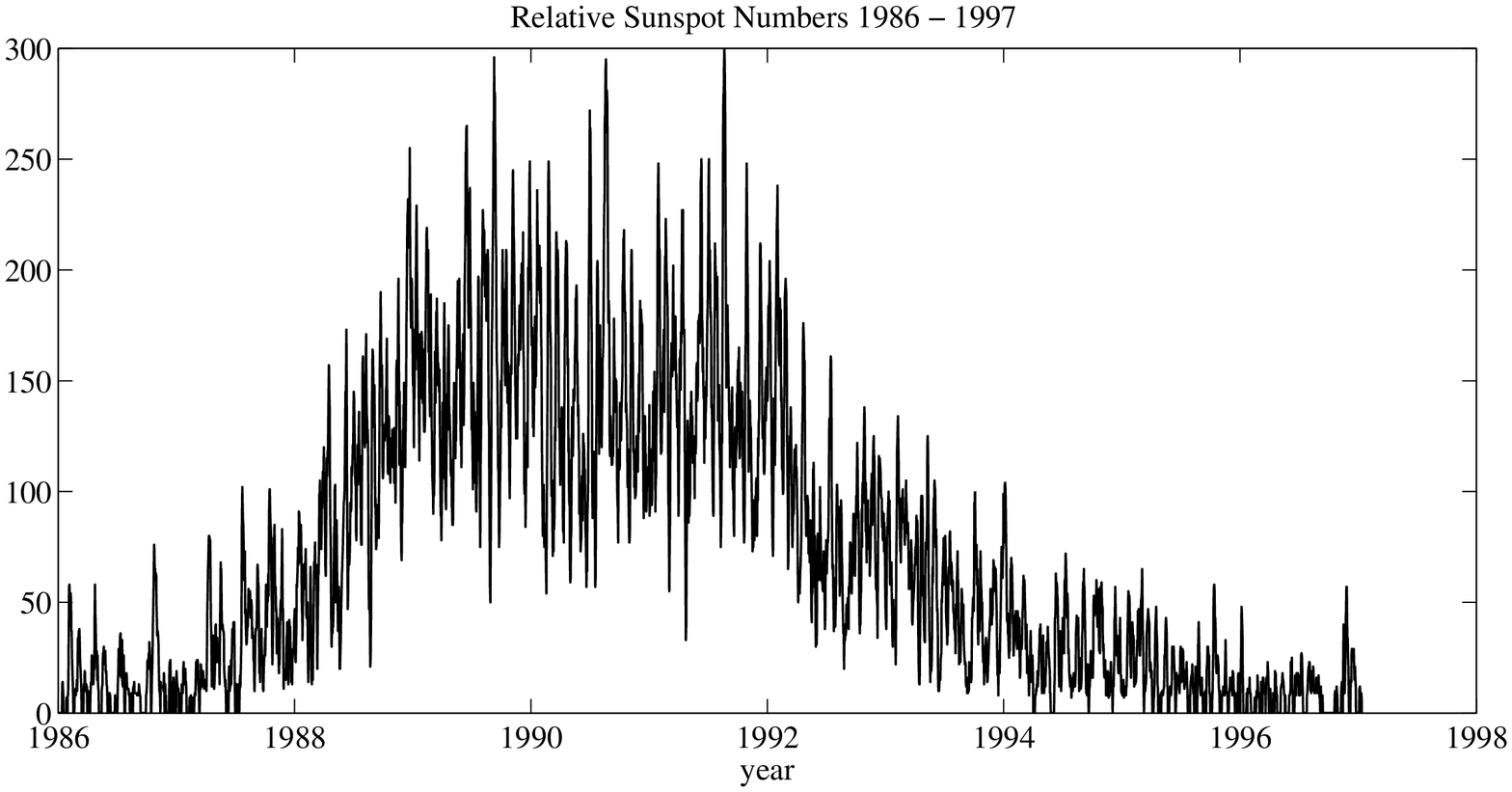}}
 \caption{Daily averaged relative sunspot numbers 1986 - 1997.}
\label{Fig9}

\centerline{
\includegraphics[width=120mm]{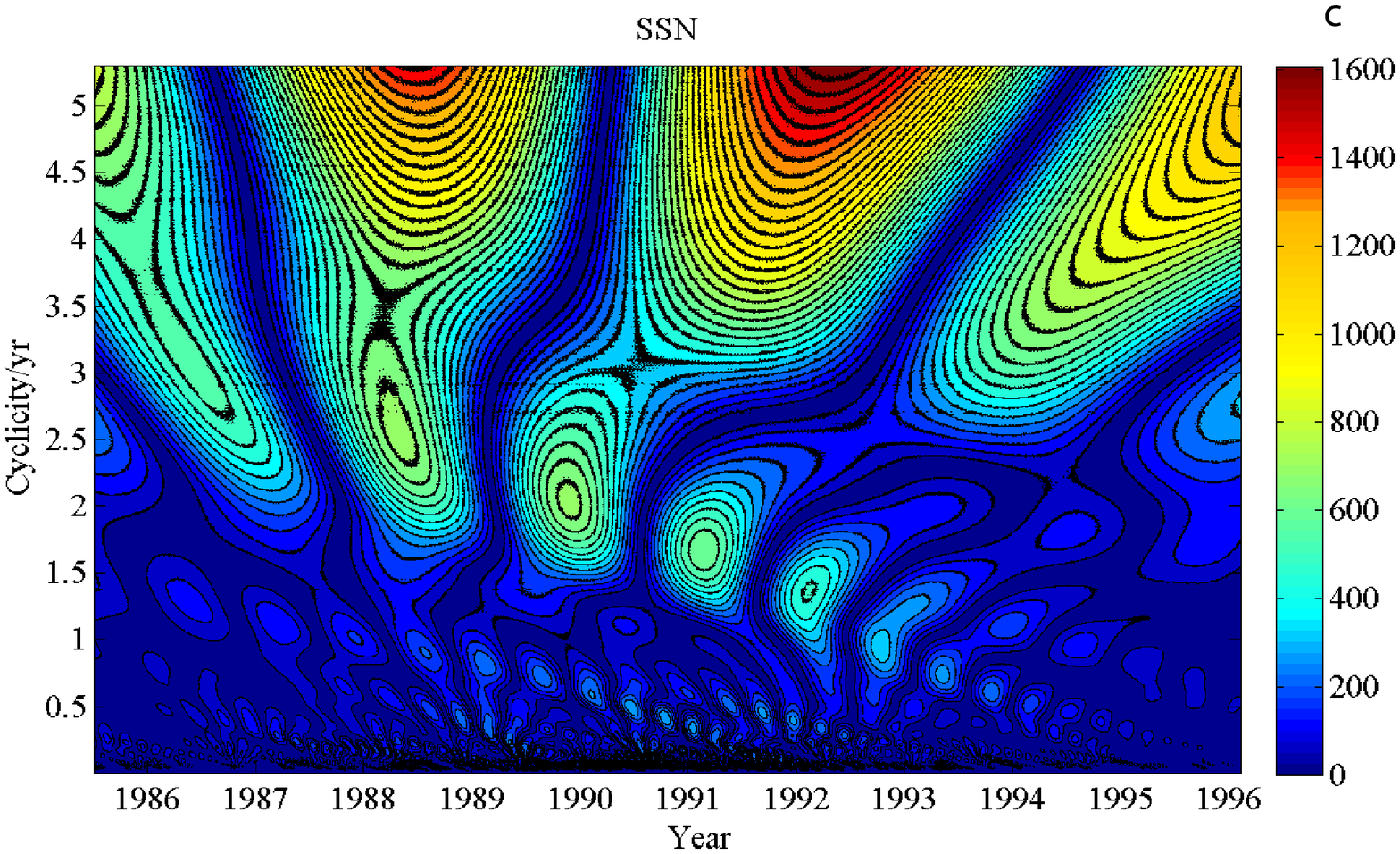}}
 \caption{The wavelet image of the cyclic activity of relative sunspot numbers on the quasi-biennial time scale.}
\label{Fig10}
\end{figure}

In Figures 6,8 we can see that the periods of cycles on different
time scales are not constant too.

Also as in the case of the learning of the Schwabe cycle we see that
approximately during three cycles value of the periods  decreases
(for the Gleissberg cycle from the periods of 110 yrs  to 70 yrs,
for the Schwabe cycle from the periods of 12 years  to 8 years -
Figures 3,6). Then during the next cycle there are two equal
amplitude cycles (two Gleissberg cycles with periods which change
from 130 to 60 yrs - Figure 3, two Schwabe cycles with periods which
change from 13 to 7 yrs -- Figures 6,8). In the following activity
cycle only the cycles with the greatest periods remain and then  the
value of the periods gradually decreases over the next three cycles.

In Figure 10 we can see that in the case of quasi-biennial cycles
the behavior of these periods inside the 11-yr cycle is similar to
the variation of cycle's periods of the Schwabe cycle and the
Gleissberg cycle. The periods of quasi-biennial cycles change from
3.5 to 2 yr inside the 11-yr cycle.

For the solar-type F,G and K stars according to {\it Kepler}
observations it was also found "shorter" chromosphere cycles with
periods of about two years, see [18],[19]. In [20] the "shorter"
cycle (like solar quasi-biennial) was determined for the star V CVn,
it's duration is equal to 2.7 yr.

\vskip12pt
\section{Observations of magnetic cycles of HK-project stars}
\vskip12pt

The most sensitive indicator of the chromospheric activity (CA) is
the Mount Wilson $S$ - index ($S_{HK}$) - the ratio of the core of
the CaII H\&K lines to the nearby continuum, see [21]. Now the CaII
H\&K emission was established as the main indicator of CA in lower
main sequence stars.

It can be noted that among the databases of observations of
Sun-like stars with known values of $S_{HK}$ the sample of stars
of the HK-project was selected most carefully in order to study
stars which are analogues of the Sun. Moreover, unlike different
Planet Search Programs of observations of Sun-like stars, the
Mount Wilson Program was specifically developed for a study of a
Sun-like cyclical activity of main sequence F, G and K-stars
(single) which are the closest to the "young Sun" and "old Sun".

The duration of observations (more than 40 years) in the HK-project
has allowed to detect and explore the cyclical activity of the
stars, similar to 11-yr cyclical activity of the Sun. First O.
Wilson began this program in 1965. He attached great importance to
the long-standing systematic observations of cycles in the stars.
Fluxes in passbands of 0.1 nm wide and centered on the CaII H\&K
emission cores have been monitored in 111 stars of the spectral type
F2-K5 on or near main sequence on the Hertzsprung-Russell diagram,
see [6],[13].

For the HK-project, stars were carefully chosen according to those
physical parameters, which are  most close to the Sun: cold,
single stars -- dwarfs, belonging to the main sequence. Close
binary systems are excluded.

Results of  joint observations of the HK-project  radiation fluxes
and periods of rotation gave the opportunity for the first time in
stellar astrophysics [22] to detect the rotational modulation of the
observed fluxes. This meant that on the surface of a star there are
inhomogeneities those are living and evolving in several periods of
rotation of the stars around its axis. In addition, the evolution of
the periods of rotation of the stars in time clearly pointed to the
fact of existence of the star's differential rotations similar to
the Sun's differential rotations.

The authors of the HK-project with use of frequency analysis of
the 40-year observations have discovered that
 the periods of 11-yr cyclic activity vary
little in size for the same star from [6],[13]. Durations of cycles
vary from 7 to 20 years for different stars. It was shown that stars
with cycles  represent about 30 \% of the total number of studied
stars.

The evolution of active regions on a star on a time scale of about
10 years determines the cyclic activity similar to the Sun.

In [6],[13] the regular chromospheric cyclical activity of
HK-project Sun-like stars were studied through the analysis of the
power spectral density with the Scargle's periodogram method [23].
It was pointed out that the detection of a periodic signal hidden by
a noise is frequently a goal in astronomical data analysis. So, in
[6] the periods of HK-project stars activity cycles similar to 11-yr
solar activity cycle were determined. The significance of the height
of the tallest peak of the periodogram was estimated by the false
alarm probability (FAP) function. Among the 50 stars with detected
cycles, only 13 stars (with the Sun), which are characterized by the
well-determined cyclic activity (the "Excellent" class), have been
found.

Unfortunately, the Scargle's periodogram method (as well as FFT
method) allows us only to define a fixed set of main frequencies
(that determines the presence of significant periodicities in the
series of observations). In the case when the values of periods
change significantly during the interval of observations, the
accuracy of determination of periods becomes worse. It is also
impossible to obtain information about the evolution of the
periodicity in time, see [24].

In [20], the different methods, such as Fast Fourier Transform,
wavelet analyze, and generalized time-frequency distributions, have
been tested and used for analyzing temporal variations in timescales
of long-term observational data which have information on the
magnetic cycles of active stars and that of the Sun. It was shown
that the application of the wavelet analysis is preferable when
studying a series of observations of the Sun and stars. Their
time-frequency analysis of multi-decadal variability of the solar
Schwabe (11-yr) and Gleissberg (century) cycles during the last 250
years showed that one cycle (Schwabe) varies between limits, while
the longer one (Gleissberg) continually increases. By analogy from
the analysis of the longer solar record, the presence of a long-term
trend may suggest an increasing or decreasing of a multi-decadal
cycle that is presently unresolved in stellar records of short
duration.

A wavelet technique has become popular as a tool for extracting a
local time-frequency information. The wavelet transform differs from
the traditional time-frequency analysis (Fourier analysis, Scargle's
periodogram method) because of its efficient ability to detect and
quantify multi-scale, non-stationary processes, see [15]. The
wavelet transform maps a one-dimensional time series $f(t)$ into the
two-dimensional plane, related to time and frequency scales.
Wavelets are the localized functions which are constructed based on
one so-called mother wavelet $\psi(t)$. The choice of wavelet is
dictated by signal or image characteristics and the nature of the
application. Understanding properties of the wavelet analysis and
synthesis, you can choose a mother wavelet function that is
optimized for your application.

The standard algorithm of wavelet analysis as applied to
astronomical observations of the Sun and stars was discussed in
detail in [15], [20]. The choice of the Daubechies and Morlet
wavelets as the best mother wavelets $\psi(t)$ for astronomical data
processing is discussed in [10],[21] on the basis of comparative
analysis of results obtained using different mother wavelets.

The results of our solar data wavelet analysis are presented on the
time-frequency plane (Figures 3,6,8,10). The notable crowding of
horizontal lines on the time-frequency plane around the specific
frequency indicates that the probability of existence of stable
cycles is higher for that frequency (cycle's duration) in accordance
with the gradient bar to the right.

In Figures 3,6,8 the results of the continuous wavelet transform
analysis (with help of Daubechies mother wavelet) of time series of
SSN are presented: Figures 3,6 corresponds to yearly averaged data,
Fig 8 corresponds to monthly averaged data, Fig 10 corresponds to
daily data. The (X, Y) plane is the time-frequency plane of
calculated wavelet-coefficients C(a, b): a-parameter corresponds to
Y plane (Cyclicity, years), b-parameter corresponds to X plane
(Time, years). The modules of C(a,b) coefficients, characterizing
the probability amplitude of regular cyclic component localization
exactly at the point (a, b), are laid along the Z axis. In Figure 5
we see the projection of C(a,b) to (a, b) or (X, Y) plane. This
projection on the plane (a, b) with isolines allows to trace changes
of the coefficients on various scales in time and reveal a picture
of local extrema of these surfaces. It is the so-called skeleton of
the structure of the analyzed process. We can also note that the
configuration of the Morlet and Daubechies wavelets are very compact
in frequency, which allows us to determine the localization of
instantaneous frequency of observed signal most accurately (compared
to other different mother wavelets).

Figures 3,6 confirms the known fact that the period of the main
solar activity cycle is about 11-yr in the XIX century and is about
10 yr in the XX century. It is also known that the abnormally long
23-rd cycle of solar activity ended in 2009 and lasted about 12.5
years. Thus, it can be argued that the value of a period of the main
cycle of solar activity for past 200 years is not constant and
varies by 10 --15 \%.

In Figure 10 we show the results of wavelet analysis of daily SSN
data in the solar cycle 22. We analyzed that data on the time scale
which is equal to several years and identified the second order
periodicity such as 5.5 years and quasi-biennial as well as their
temporal evolution.

In [25] a study of time variations of cycles of 20 active stars
based on decades (long photometric or spectroscopic observations)
with a method of time-frequency analysis was done. They found that
cycles of sun-like stars show systematic changes. The same
phenomenon can be observed for the cycles of the Sun.

In [25] was found that fifteen stars definitely show multiple
cycles, the records of the rest are too short to verify a timescale
for a second cycle. For 6 HK-project stars (HD 131156A, HD 131156B,
HD 100180, HD 201092, HD 201091 and HD 95735) the multiple cycles
were detected. Using wavelet analysis the following results (other
than periodograms from [6]) were obtained:

HD 131156A shows variability on two time scales: the shorter cycle
is about 5.5-yr, a longer-period variability is about 11 yr.

For HD 131156B only one long-term periodicity has been determined.

For HD 100180 the variable cycle of 13.7-yr appears in the beginning
of the record; the period decreases to 8.6-yr by the end of the
record. The results in the beginning of the dataset are similar to
those found by [6], who found two cycles, which are equal to 3.56
and 12.9-yr.

The record for HD 201092 also exhibits two activity cycles: one is
equal to 4.7-yr, the other has a time scale of 10-13 years.

The main cycle, seen in the record of HD 201091, has a mean length
of 6.7-yr, which slowly changes between 6.2 and 7.2-yr. A shorter,
significant cycle is found in the first half of the record with a
characteristic time scale of 3.6-yr.

The stronger cycle of HD 95735 is 3.9-yr. A longer, 11-yr cycle is
also present with a smaller amplitude.

In our paper we have applied the wavelet analysis for partially
available data from the records of relative CaII emission fluxes -
the variation of $S_{HK}$ for 1965-1992 observation sets from [6]
and for 1985-2002 observations from [13].

We used the detailed plots of $S_{HK}$ time dependencies: each point
of the record of observations, which we processed in this paper
using wavelet analysis technique, corresponds to three months
averaged values of $S_{HK}$.

In this paper we have studied 5 HK-project stars with cyclic
activity of the "Excellent" class: HD 10476,  HD 81809, HD 103095,
HD 152391, HD 160346 and the star HD 185144 with no cyclicity
(according [6] classification).

We used the Daubechies 10 wavelet which can most accurately
determine the dominant cyclicity as well as its evolution in time in
solar data sets at different wavelengths and spectral intervals, see
[3].

We hope that wavelet analysis can help to study the temporal
evolution of  chromospheric cycles of the stars. Tree-month
averaging also helps us to avoid the modulation of observational
$S_{HK}$ data by star's rotations.

In Figures 11-22 we present our results for cycles of 6 Mount Wilson
HK-project stars.

Figures 11,12 show 3-monthly averaged HD 10476 data set from [6] and
[13] plots and the wavelet image of the cyclic activity of HD 10476.
We can see that HD 10476 has a mean cycle duration of 10.0-yr in the
first half of the record, then it sharply changes to 14-yr, while in
[6] was found duration of 9.6-yr. After changing the high amplitude
cycle's period from 10-yr to 14-yr in 1987, the low amplitude cycle
remained with 10.0-yr period -- we can see two activity cycles. In
[6] the duration of HD 10476 mean cycle estimated as 9.6-yr.

Figures 13,14 show 3-monthly averaged HD 81809 data set from [6] and
[13] plots and the wavelet image of the cyclic activity of HD 81809.
We can see that HD 81809 has a mean cycle duration of 8.2-yr, which
slowly changes between 8.3-yr in the first half of the record and
8.1-yr in the middle and the end of the record while in [6] found
8.17-yr.

Figures 15,16 show 3-monthly averaged HD 103095 data set from [6]
and [13] plots and the wavelet image of the cyclic activity of HD
103095. HD 103095 has a mean cycle duration of 7.2-yr, which slowly
changes between 7.3-yr in the first half of the record, 7.0-yr in
the middle and 7.2-yr in the end of the record while in [6] found
7.3-yr.

Figures 17,18 show 3-monthly averaged HD 152391 data set from [6]
and [13] plots and the wavelet image of the cyclic activity of HD
152391. HD 152391 has a mean cycle duration of 10.8-yr, which slowly
changes between 11.0-yr in the first half of the record and 10.0-yr
in the end of the record while in [6] found 10.9-yr.

Figures 19,20 show 3-monthly averaged HD 160346 data set from [6]
and [13] plots and the wavelet image of the cyclic activity of HD
160346. HD 160346 has a mean cycle duration of 7.0-yr which does not
change during the record in agreement with [6] estimated 7.0-yr.

Figures 21,22 show 3-monthly averaged HD 185144 data set from [6]
and [13] plots and the wavelet image of the cyclic activity of HD
185144. HD 185144 has a mean cycle duration of 7-yr which changes
between 8-yr in the first half of the record and 6-yr in the end of
the record while in [6] haven't found the well-pronounced cycle.

In [25] the multiple cycles were found for HD 13115A, HD 131156B, HD
93735 stars, for which no cycles have been found in [6]. For the
stars of the "Excellent" class HD 201091 and HD 201092, cycle
periods found in[6] were confirmed and the shorter cycles (similar
to solar quasi-biennial) were also determined.

In [25] have concluded that all the stars from their pattern of cool
main sequence stars have cycles and most of the cycle durations
change systematically.

However we can see that the stars of the "Excellent" class have
relatively constant cycle durations~-- for these stars the cycle's
periods calculated in [6] and cycle's periods found with the use of
the wavelet analysis are the same.

\begin{figure}[tbh!]
\centerline{
\includegraphics[width=80mm]{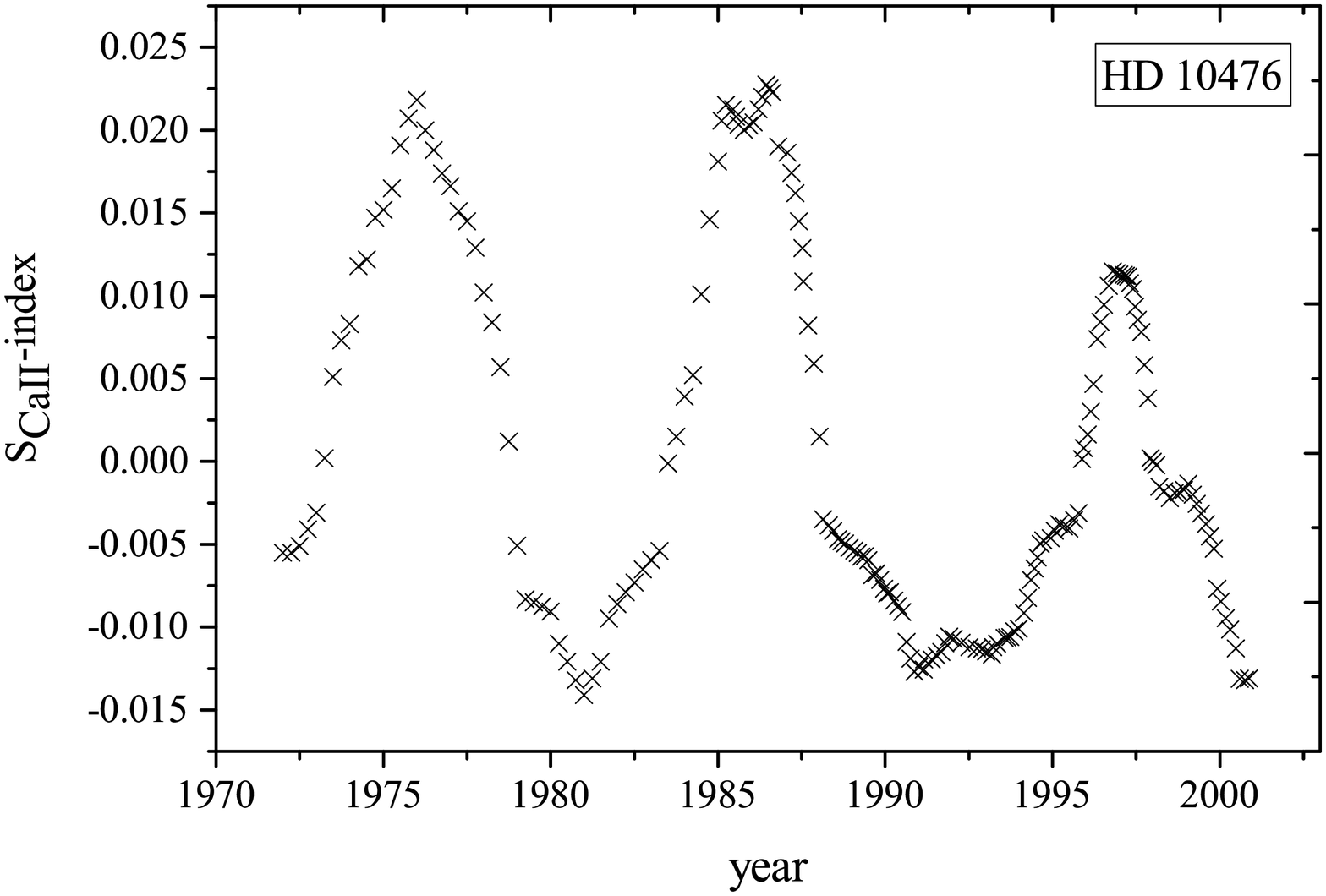}}
 \caption{3-monthly averaged observations of HD 10476 from 1968 to 2002.}
{\label{Fi:Fig11}} \centerline{
\includegraphics[width=120mm]{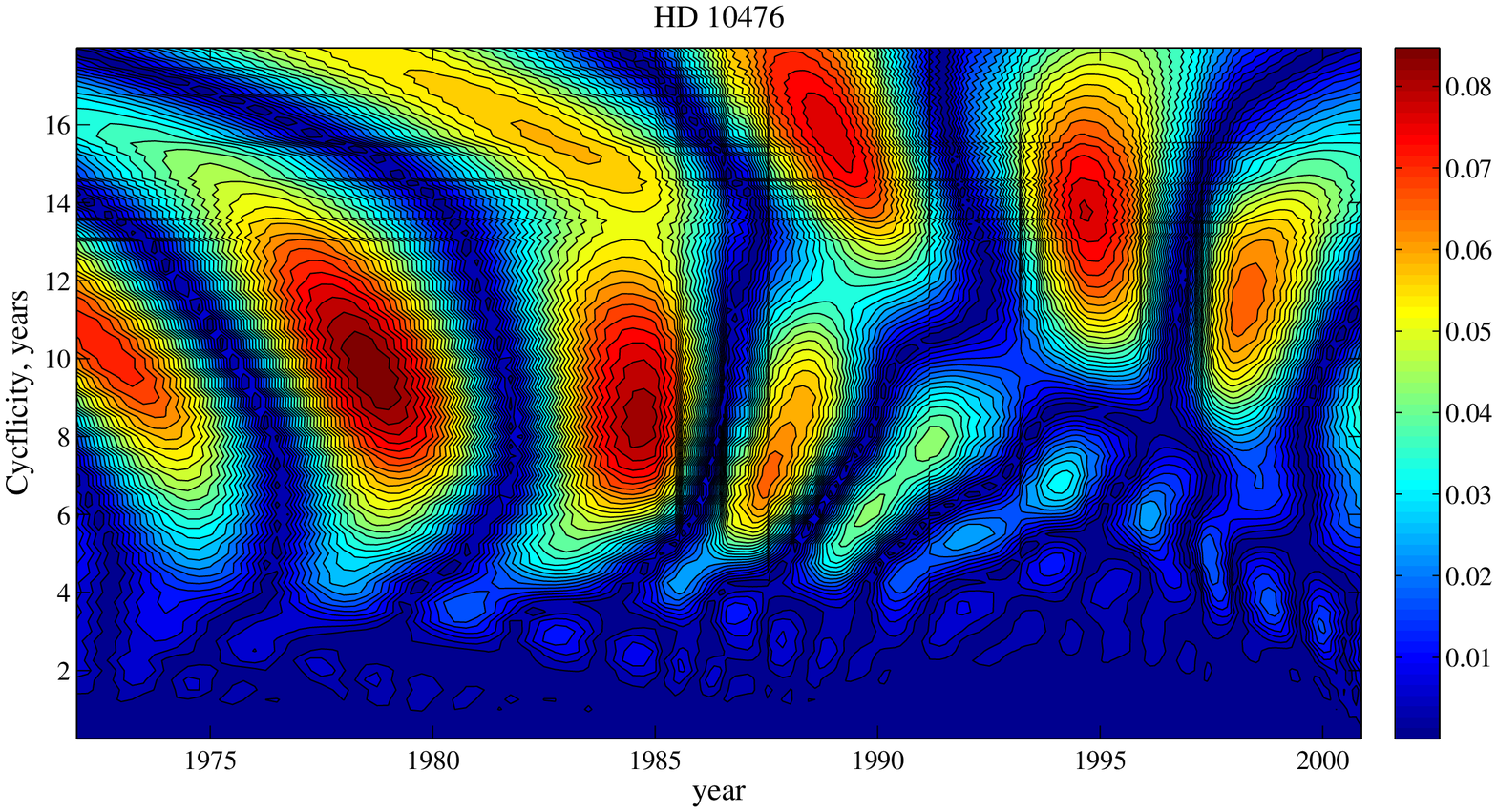}}
 \caption{The wavelet image of the cyclic activity of HD 10476}
\label{Fig12}
\end{figure}

\begin{figure}[tbh!]
\centerline{
\includegraphics[width=80mm]{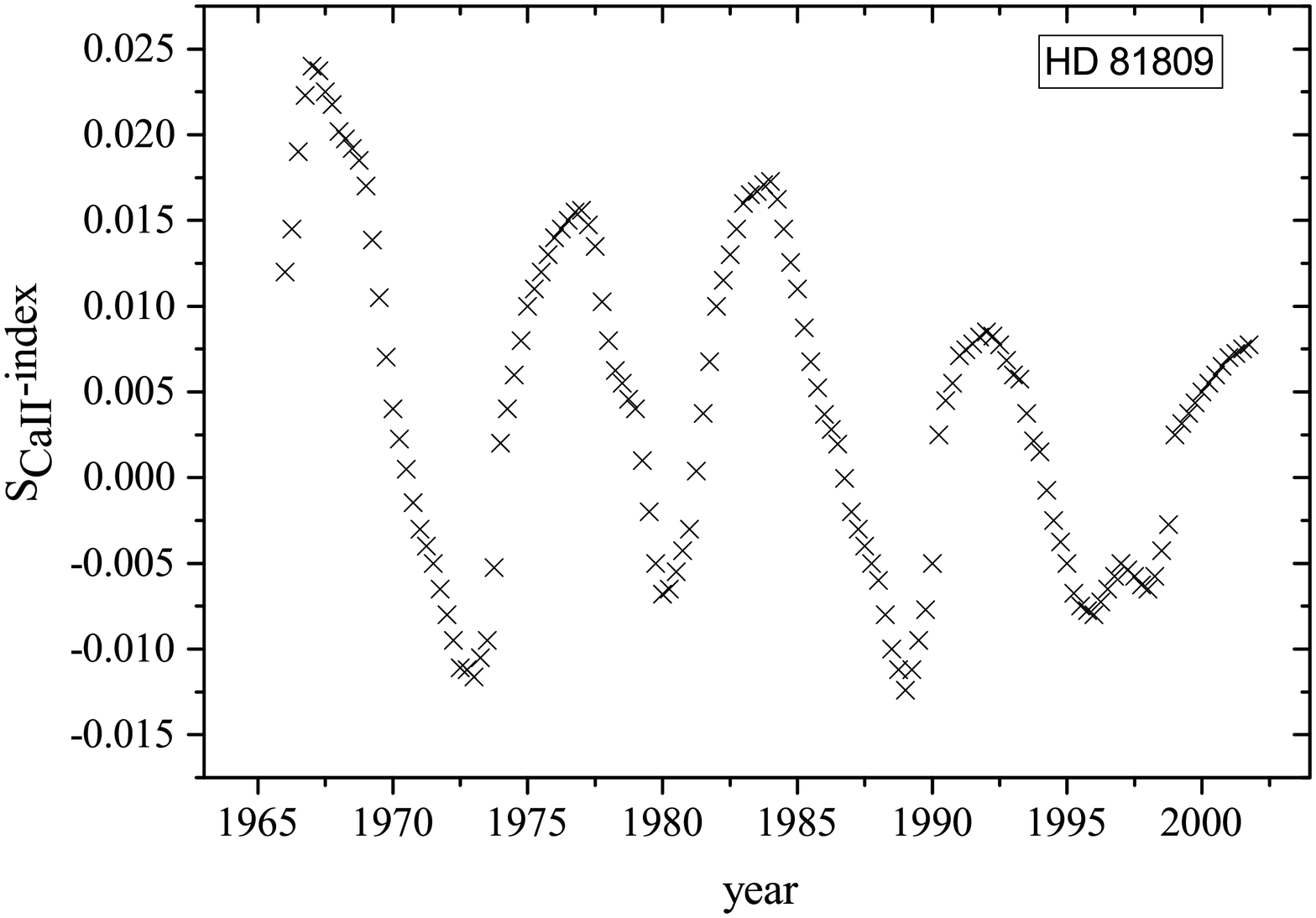}}
 \caption{3-monthly averaged observations of HD 81809 from 1968 to 2002.}
\label{Fig13}

\centerline{
\includegraphics[width=120mm]{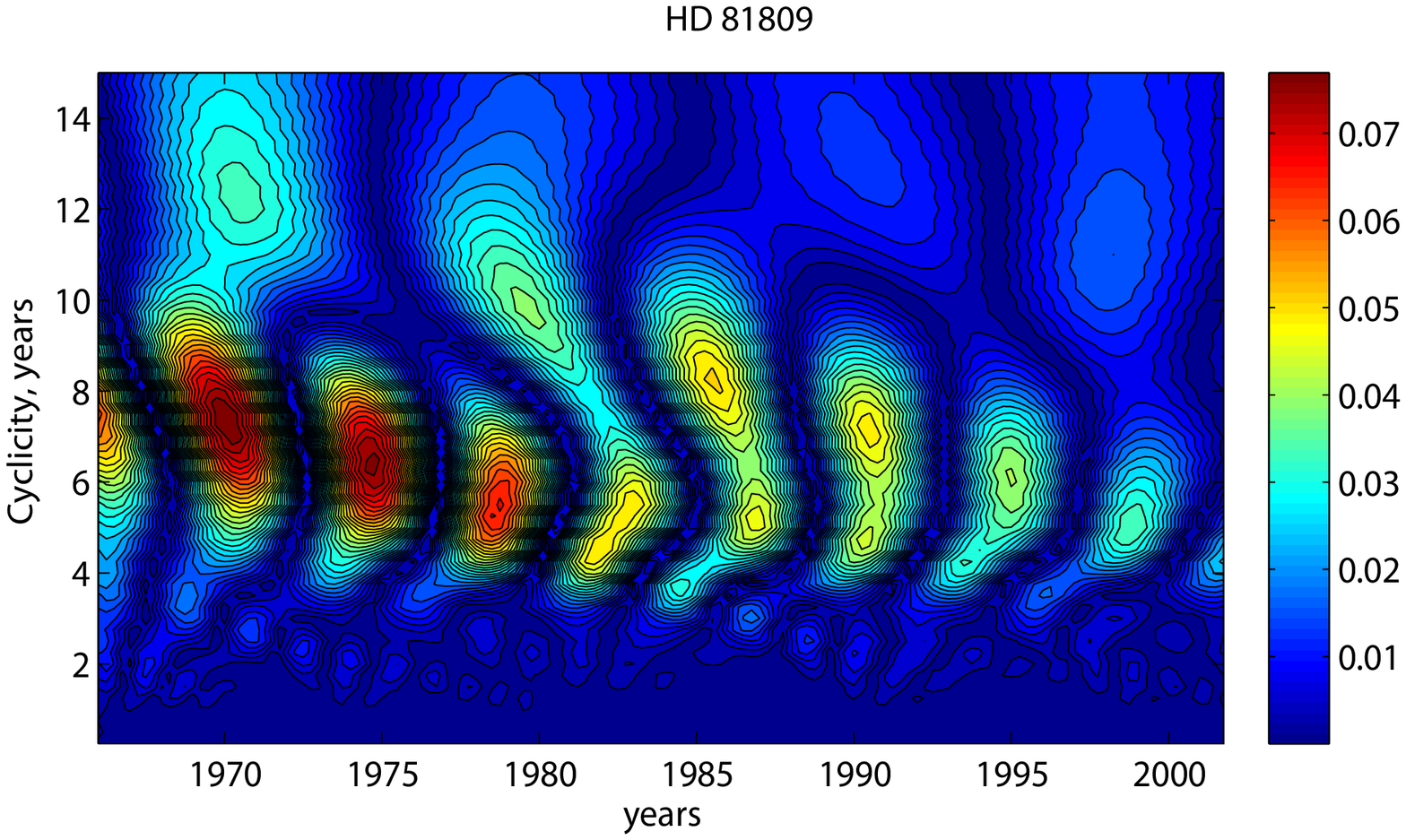}}
 \caption{The wavelet image of the cyclic activity of HD 81809}
\label{Fig14}
\end{figure}

\begin{figure}[tbh!]
\centerline{
\includegraphics[width=80mm]{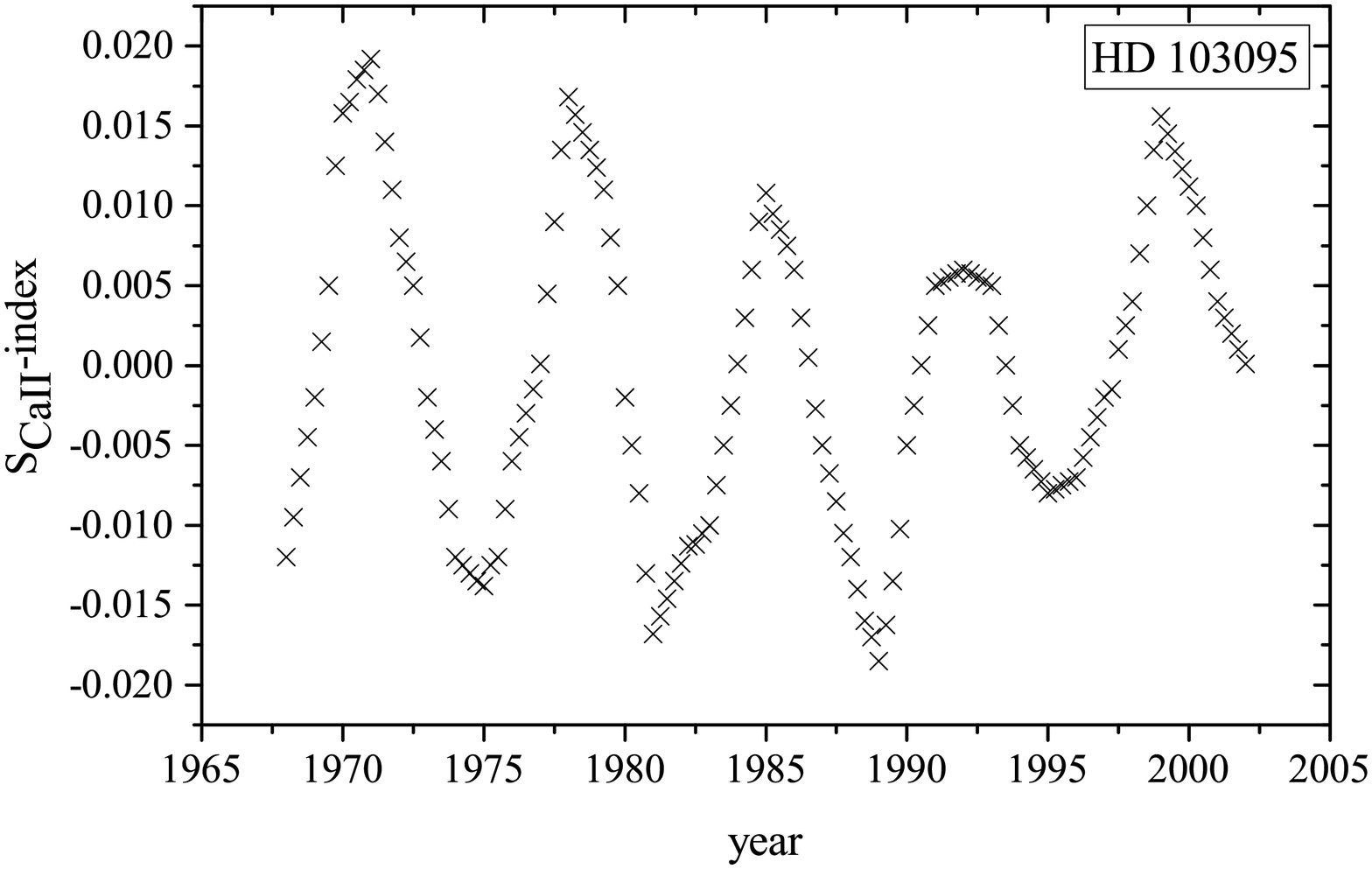}}
 \caption{3-monthly averaged observations of HD 103095 from 1968 to 2002.}
\label{Fig15}

\centerline{
\includegraphics[width=120mm]{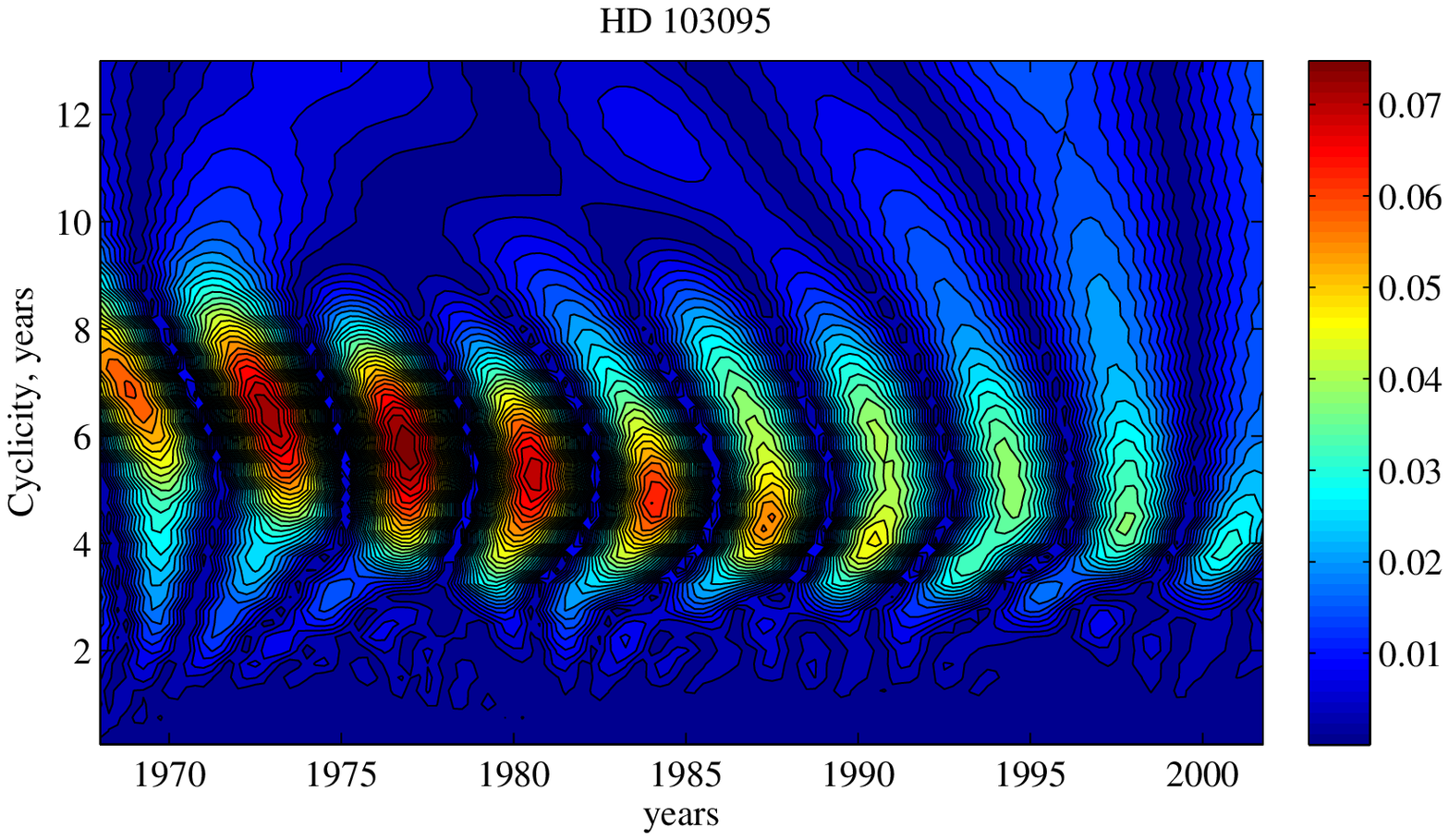}}
 \caption{The wavelet image of the cyclic activity of HD 103095.}
\label{Fig16}
\end{figure}

\begin{figure}[tbh!]
\centerline{
\includegraphics[width=80mm]{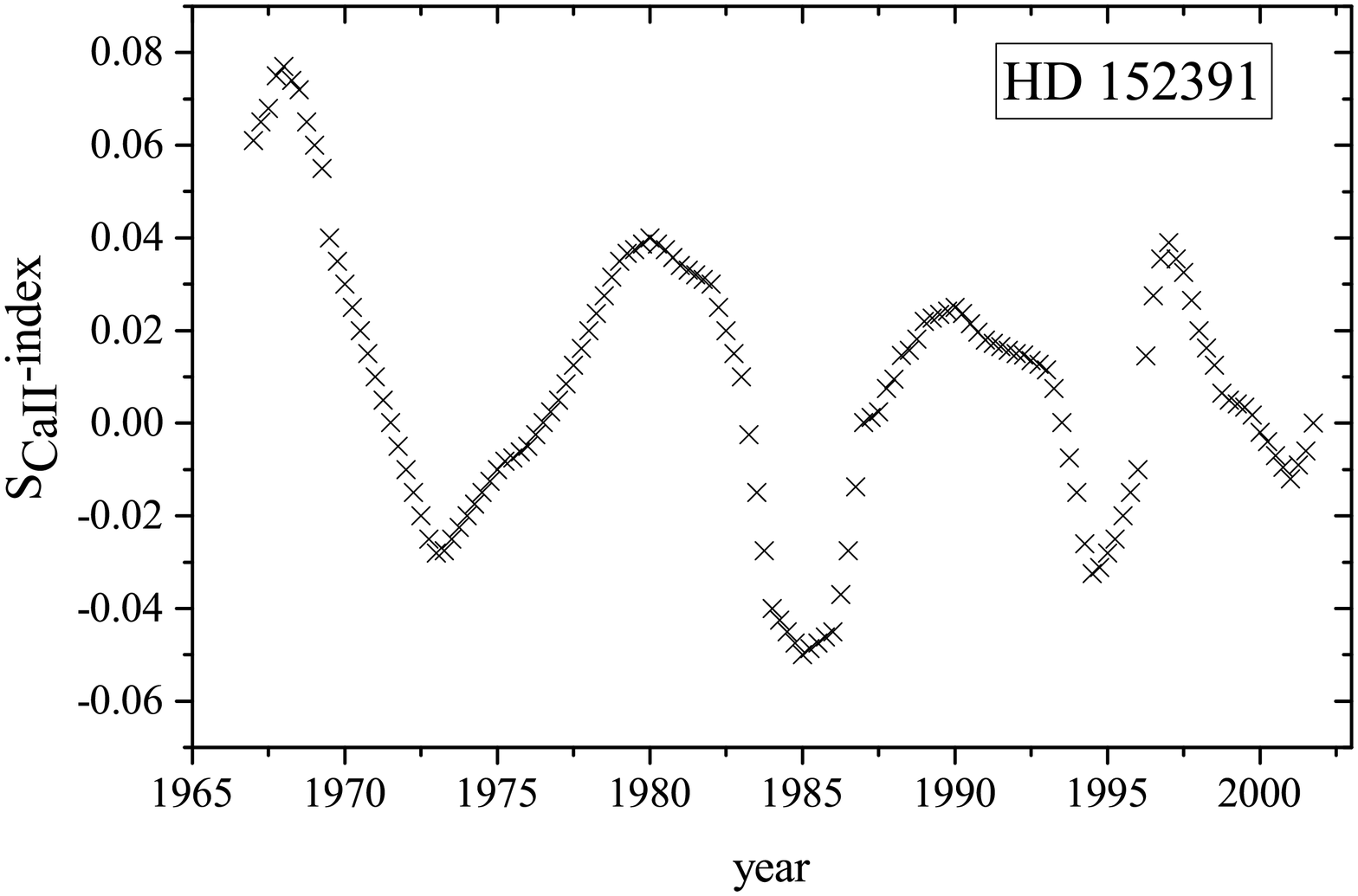}}
 \caption{3-monthly averaged observations of HD 152391 from 1966 to 2002.}
\label{Fig17}

\centerline{
\includegraphics[width=120mm]{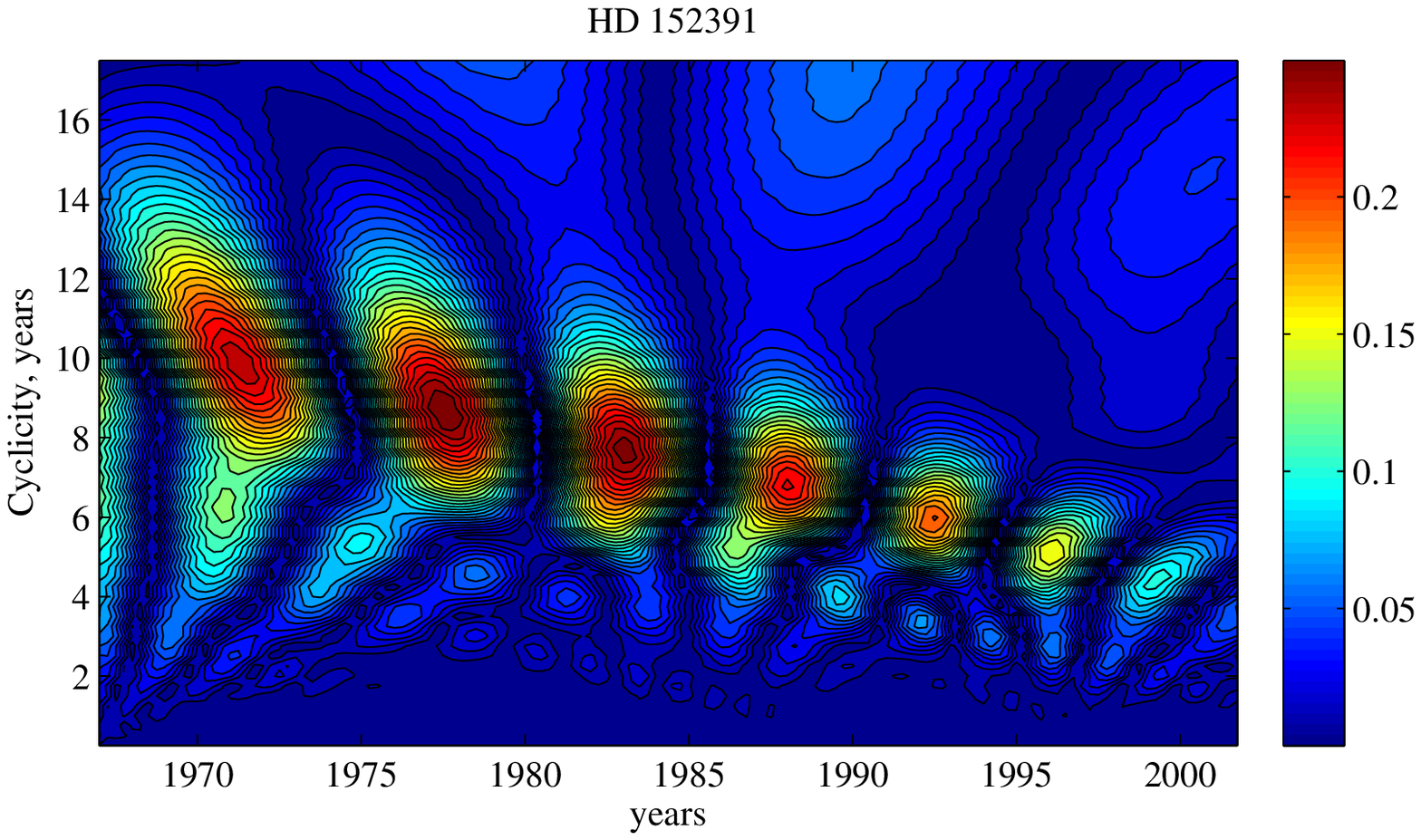}}
 \caption{The wavelet image of the cyclic activity of HD 152391.}
\label{Fig18}
\end{figure}

\begin{figure}[tbh!]
\centerline{
\includegraphics[width=80mm]{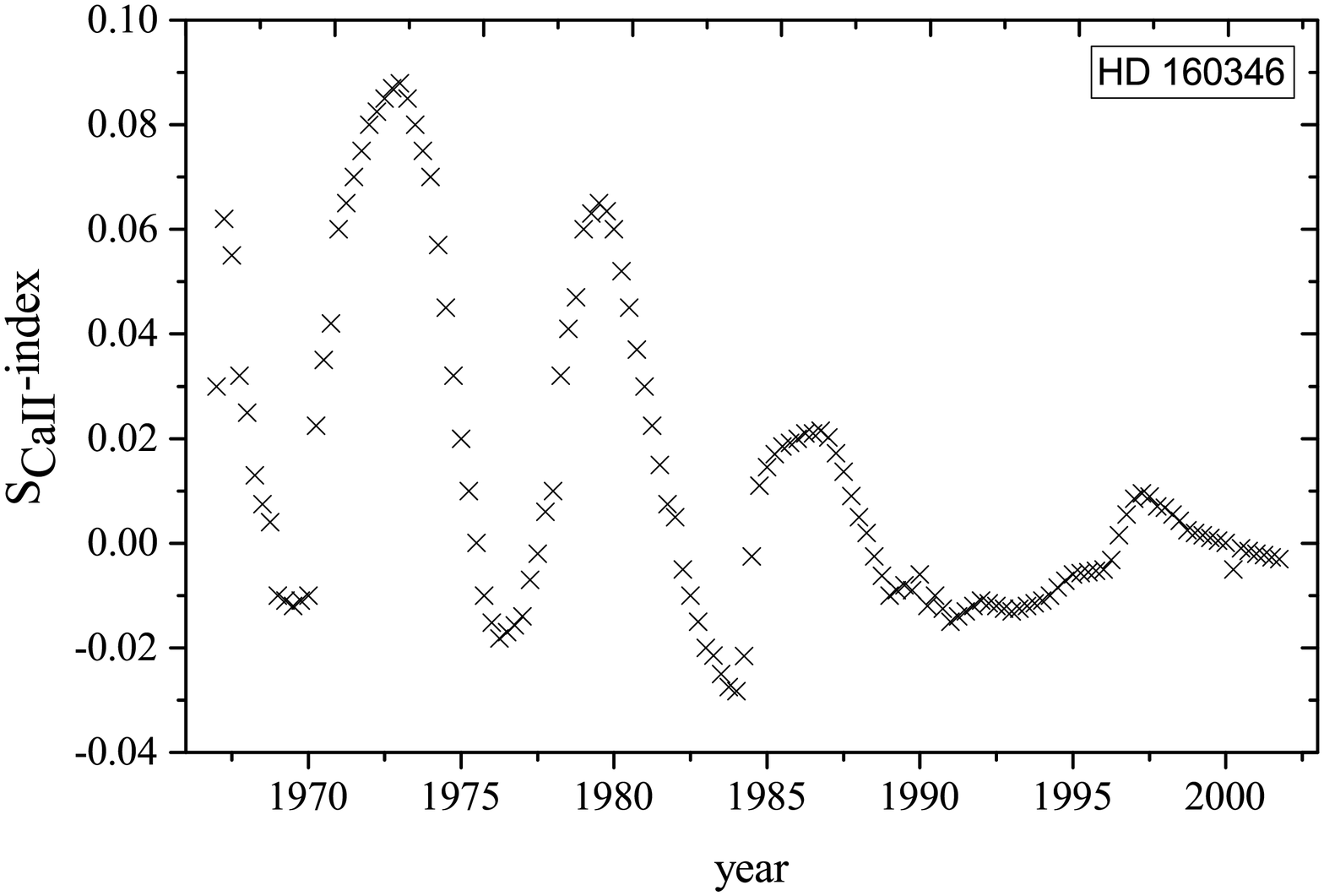}}
 \caption{3-monthly averaged observations of HD 160346 from 1966 to 2002.}
\label{Fig19}

\centerline{
\includegraphics[width=120mm]{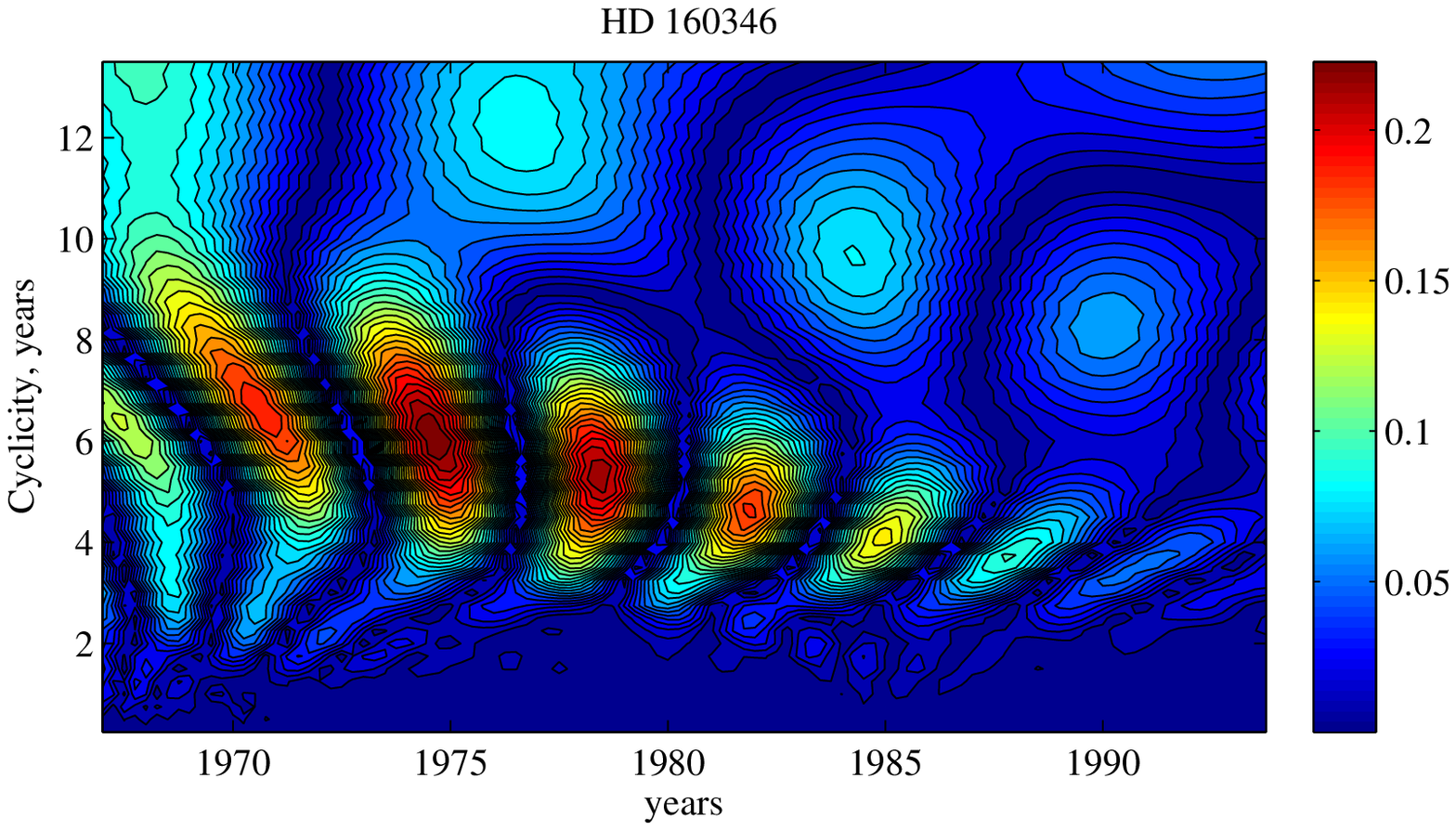}}
 \caption{The wavelet image of the cyclic activity of HD 160346.}
\label{Fig20}
\end{figure}

\begin{figure}[tbh!]
\centerline{
\includegraphics[width=80mm]{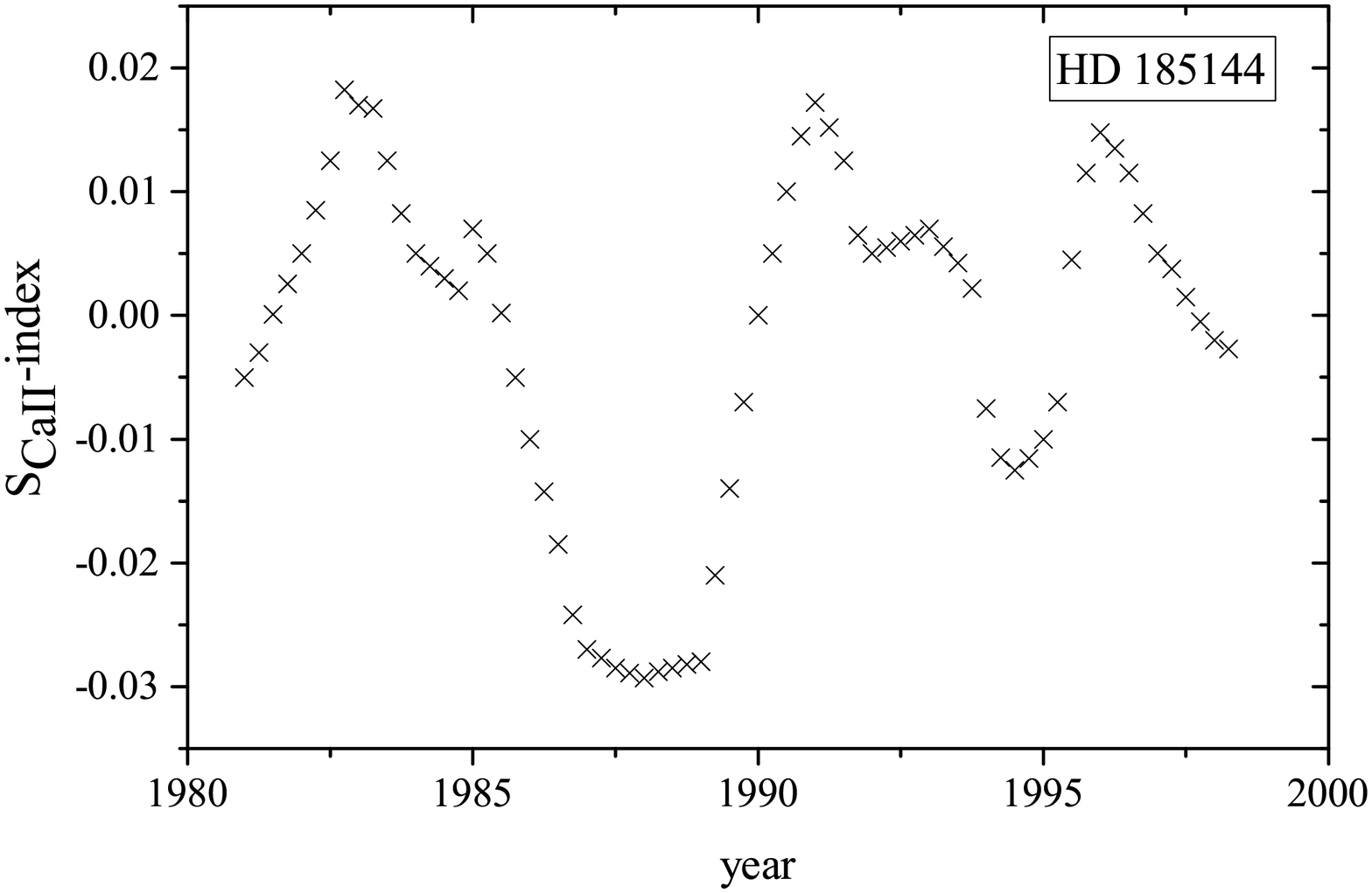}}
 \caption{3-monthly averaged observations of HD 185144 from 1981 to 1999.}
\label{Fig21}

\centerline{
\includegraphics[width=120mm]{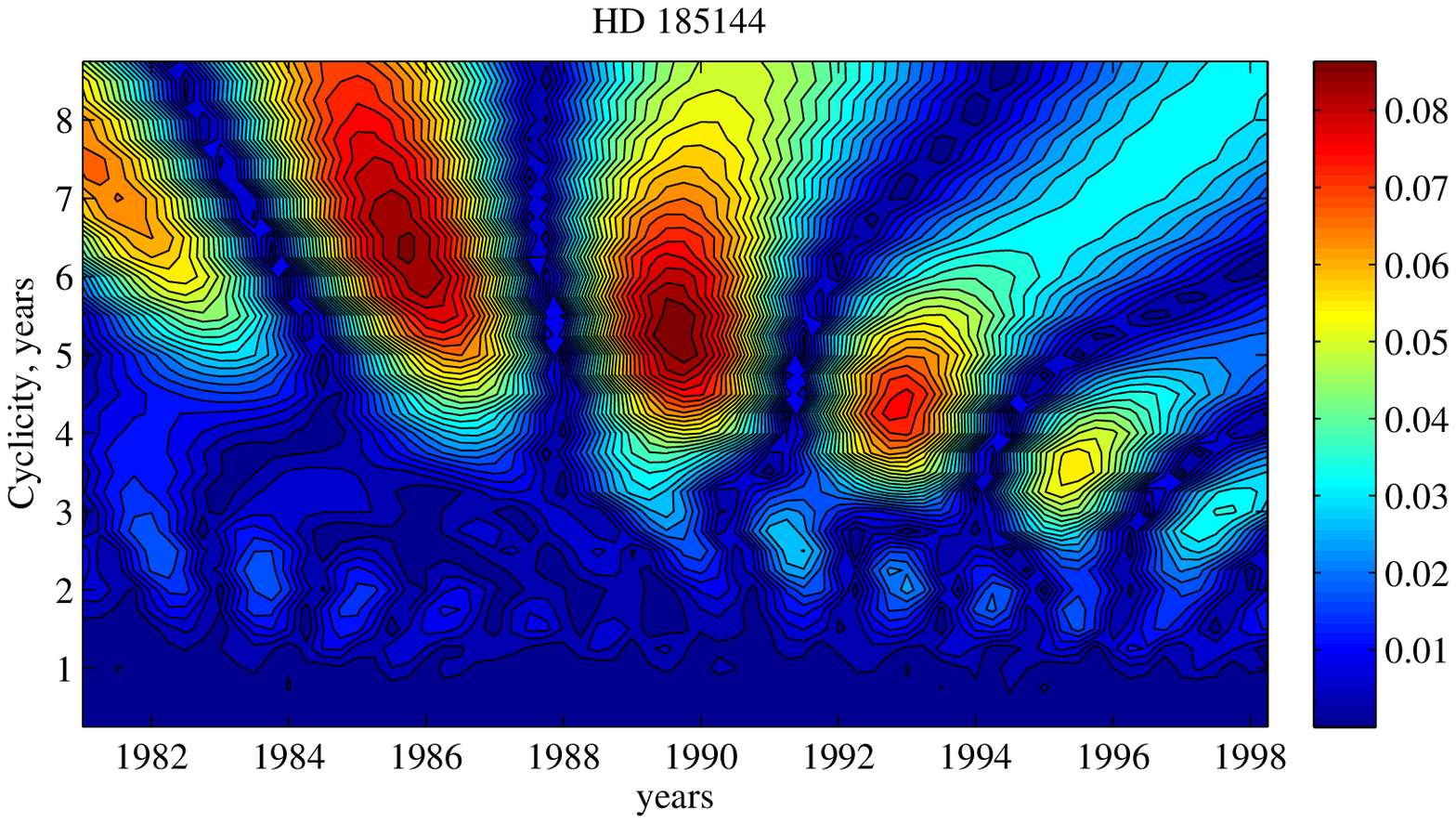}}
 \caption{The wavelet image of the cyclic activity of HD 185144.}
\label{Fig22}
\end{figure}

A similar picture can be seen for the Sun: the long-term behavior of
the sunspot group numbers has been analyzed using a wavelet
technique by [15] who plotted changes of the Schwabe cycle (length
and strength) and studied the grand minima. The temporal evolution
of the Gleissberg cycle can also be seen in the time-frequency
distribution of the solar data. According to [15], the Gleissberg
cycle is as variable as the Schwabe cycle. It has two higher
amplitude occurrences: the first one is around 1800 (during the
Dalton minimum), and the next one is around 1950. They found very
interesting fact~-- the continuous decrease in the frequency
(increase of period) of the Gleissberg cycle. While near 1750 the
cycle duration was about 50 yr, it lengthened to approximately 130
yr by 1950.

In the late part of the XX century, some of solar physicists began
to examine with different methods the variations of relative sunspot
numbers not only in the high amplitude 11-yr Schwabe cycle but in
low amplitude cycles approximately equal to half (5.5-yr) and fourth
(quasi-biennial) parts of the period of the main 11-yr cycle, see
[16]. The periods of the quasi-biennial cycles vary considerably
within one 11-yr cycle, decreasing from 3.5 to 2 yrs (see Figure
8,10), which complicates a study of such periodicity with the
periodogram method.

Using methods of frequency analysis of signals, the quasi-biennial
cycles have been studied not only for the relative sunspot number,
but also for 10.7 cm solar radio emission and for some other indices
of solar activity, see [3]. It was also shown that the cyclicity on
the quasi-biennial time scale often takes place among stars with
11-yr cyclicity, see [10,14].

The cyclicity similar to the solar quasi-biennial was also detected
for Sun-like stars from direct observations. In [26], the results of
direct observations of magnetic cycles of 19 Sun-like stars of F, G,
K spectral classes within 4 years were presented. Stars of this
sample are characterized by masses between 0.6 and 1.4 of the solar
mass and by rotation periods between 3.4 and 43 days. Observations
were made using NARVAL spectropolarimeter (Pic du Midi, France)
between 2007 and 2011. It was shown that for the stars of this
sample $\tau$ Boo and HD 78366 (the same of the Mount Wilson
HK-project) the cycle lengths derived by CA
 by [6] seem to be longer than those
derived by spectropolarimetry observations of [26]. They suggest
that this apparent discrepancy may be due to the different temporal
sampling inherent to these two approaches, so that the sampling
adopted at Mount Wilson may not be sufficiently tight to unveil
short activity cycles. They hope that future observations of Pic du
Midi stellar sample will allow them to investigate longer time
scales of the stellar magnetic evolution.

For the Sun-like F, G and K stars according to {\it Kepler}
observations, "shorter" chromosphere cycles with periods of about
two years have also been found, see [18],[19].

We assume that precisely these quasi-biennial cycles were identified
in [26]: $\tau$ Boo and HD 78366 are the same of the HK-project,
these stars have cycles similar to the quasi-biennial solar cycles
with periods of a quarter of the duration of the periods defined in
[6].

Note, that in case of the Sun, the amplitude of variations of the
radiation in quasi-biennial cycles is substantially less than the
amplitude  of variations in main 11-yr cycles. We believe that
this fact is also true for all Sun-like stars of the HK-project
and in the same way for $\tau$ Boo and HD 78366.

The quasi-biennial cycles cannot be detected with the Scargle's
periodogram method. But methods of spectropolarimetry from [26]
allowed detecting the cycles with 2 and 3-yr periods. Thus,
spectropolarimetry is more accurate method for detection of cycles
with different periods and with low amplitudes of variations.

So, the need for wavelet analysis of HK-project observational data
is dictated also by the fact that the application of the wavelet
method to these observations will help: (1) to find cyclicities
with periods equal to a half and a quarter from the main high
amplitude cyclicity; (2) to clarify periods of high amplitude
cycles and to follow their evolution in time; (3) to find other
stars with cycles for which cycles were not determined using the
periodogram method due to strong variations of the period as in
the case of HD 185144.

\vskip12pt
\section{The parameters of time-evolution study of the cycles}
\vskip12pt

To describe this general trend we propose a formal representation of
this process. The cyclic variations of fluxes of solar radiation (in
particular, the SSN as the most frequently studied activity index)
can be represented by a sinusoid with varying period and constant
amplitude:

$$ A(t) = cos (2\pi\frac{t-t_0}{T}) $$

Note that exactly this behavior we can see in different cycles of
activity, see Fig.2, Fig.4, Fig.5.

The smooth change of the cycle period can be represented as follows:

$$ T(t)=T_0 - k(t) \cdot (t-t_0) $$

where $~t_0$ is the peak time of the studied cycle, $T_0$ is the
cycle's period at the time $t_0$, t varies in the range $t_0<t<
t_0+T_0$.

\begin{table}[tbh!]
\caption{Parameters of different solar cycles.} \vskip12pt

\centering
\begin{tabular}{clclcl}

\hline \hline
&&&      \\
~~~~Cyclicity~~~&~~~Cycle's period~~~&~~~ k(t)  \\
\hline
&&&      \\
~~~Solar SSN Century cycle~~~&~~~~~100 yr~~~&~~~~0.3~\\
\hline
&&&      \\
~~~Solar A(mvh) Two-Century cycle~~~&~~~~~200 yr~~~&~~~~0.3~\\
\hline
&&&      \\
~~~Solar SSN Half a century cycle~&~~~~~ 50 yr~~~&~~~~~0.25~\\
\hline
&&&      \\
~~~Solar SSN 11-yr cycle~~~ &~~~~~10~-11 yr~~ &~~~~ 0.2~  \\
\hline
&&&      \\
~~~Solar SSN Quasi-biennial cycle~~~&~~~~~ 2 - 3.5 yr~~~&~~~~~0.33~\\
\hline
&&&      \\
~~~Star's HD 81809 cycle~~~&~~~~~ 6 - 8 yr~~~&~~~~~0.25~\\
\hline
&&&      \\
~~~Star's HD 103095 cycle~&~~~~~ 5 -- 7 yr~~~&~~~~~0.33~\\
\hline
&&&      \\
~~~Star's HD 10476 cycle ~&~~~~~10 -- 14 yr~~~&~~~~0.3~\\
\hline
&&&      \\
~~~Star's HD 152391 cycle~&~~~~~ 8 -- 11 yr~~~&~~~~~0.3~\\
&&&      \\
~~~Star's HD 160346 cycle~~~ &~~~~~5 -- 7 yr~~ &~~~~ 0.33~  \\
\hline
&&&      \\
~~~Star's HD 185144 cycle~&~~~~~ 5 -- 7 yr~~~&~~~~~0.33~\\
  \hline\hline
\end{tabular}
\end{table}

In Table 1. we presented the values of k(t) for different solar
cycles. For each solar cycle (from the quasi-biennial duration to
11-yr and 200-yr cycle's periods) and also for 6 star's cycles the
values of coefficient k(t) are different, see Figures
3,6,8,10,12,14,16,18,20,22. We consider that it is necessary to take
into account the temporal evolution of solar and Sun-like star
cycles for successful forecasts or the parameters of activity
cycles.

\vskip12pt
 \vskip12pt
\section{Conclusions}
\vskip12pt

The study of the evolution of solar cyclicity by observations of the
Relative Sunspot Number and Sunspot Areas variation using the
wavelet analysis allows us to make more accurate predictions of
indices of solar activity (and consequently the predictions of the
parameters of the earth's atmosphere), and also to take a step
towards a greater understanding of the nature of cyclicity of solar
activity. The close interconnection between activity indices make
possible new capabilities in the solar activity indices forecasts.
For a long time the scientists were interested in the simulation of
processes in the earth's ionosphere and upper atmosphere. For these
purposes it is necessary the successful forecasts of maximum values
and other parameters of future activity cycles and also it has been
required to take into account the century component.

The study of the evolution of Sun-like stars cyclicity by example of
Mount Wilson observations of $S_{HK}$ - index using the wavelet
analysis reveals the similar features in solar and stellar cyclic
activity: the existence of multiple cycles and their evolution in
time.

Wavelet analysis of these data reveals the following features: the
period and phase of these relatively low frequency variations of the
solar or stellar fluxes, previous to the studied time point,
influence to the amplitudes and to the phase of studied time point.
 Solar or stellar fluxes show the gradually changing of their
values in time: as a result, the periods of variations are getting
longer.


\begin{thebibliography}{99}


\bibitem{Svalgaard11}
Svalgaard, L., Lockwood, M., Beer, J. 2011, {\it{
http://www.leif.org/research/Svalgaard\_ISSI\_Proposal\_Base.pdf}}.

\bibitem{Svalgaard10}
Svalgaard, L., Cliver, E.W. 2010, {\it {J. Geophys. Res.}},
{\bf{115}}, A09111.

\bibitem{Bruevich15}
 Bruevich, E.A. and Yakunina, G.V. 2015, {\it{
    Moscow University Physics Bulletin}}, {\bf{70}}, Issue 4, 282.


\bibitem{Bruevich07}
Bruevich, E.A., Alekseev, I.Yu. 2007, {\it{Astrophysics}},
{\bf{50}}, Issue 2, 187.

\bibitem{Bruevich01}
 Bruevich, E.A., Katsova, M.M. and Sokolov, D.D. 2001, {\it{Astronomy Reports}},
 {\bf{45}}, Issue 9, 716.

\bibitem{ Baliunas95}
Baliunas, S.L., Donahue, R.A. et al. 1995, {\it{ApJ}}, {\bf{438}},
269.

\bibitem{Wright04} Wright, J.T., Marcy, G.W., Butler, R.P., Vogt, S.S., 2004. {\it{ApJS}}, {\bf{152}}, Issue 2,
261.

\bibitem{Arriagada11} Arriagada, P. 2011, {\it{ApJ}}, 734, 70.

\bibitem{Bruevich16}
Bruevich, E.A., Bruevich, V.V., Shimanovskaya, E.V. 2016,
{\it{Astrophysics}}, {\bf{59}}, Issue 1, 101.

\bibitem{Shimanovskaya16}
Shimanovskaya, E.V, Bruevich, V.V., Bruevich, E.A. 2016,
{\it{Research in Astronomy and Astrophysics}}, {\bf{16}}, Issue 9,
148.

\bibitem{Nagovitsyn16}
Nagovitsyn, Yu.A., Tlatov, A.G.,
Nagovitsyna E.Yu. 2016, {\it{Astronomy Reports}}, {\bf{60}}, Issue
9, 831.

\bibitem{NGDC16}National Geophysical Data Center Solar and Terrestrial Physics.
 2016, http://www.ngdc.noaa.gov/stp/solar/solardataservices.html.

\bibitem{ Lockwood07}
Lockwood, G.W., Skif, B.A., Radick, R.R., Baliunas, S.L.,
    Donahue, R.A. and Soon, W. 2007,  {\it{ApJS}}, {\bf{171}}, 260.


\bibitem{Bruevich11}
 Bruevich, E.A. and Kononovich, E.V. 2011, {\it {Moscow University Physics Bulletin}}, {\bf{66}}, Issue 1, 72.

\bibitem{Frick97}
Frick,  P., Baliunas, S.L., Galyagin, D., Sokoloff, D. and Soon, W.
1997, {\it{ApJ}}, {\bf{483}}, 426.

\bibitem{Vitinsky86}
Vitinsky, Yu., Kopezky, M., Kuklin G., 1986. The sunspot solar
activity statistic. M. Nauka

\bibitem{Bruevich14a} Bruevich, E.A.,
Bruevich, V.V., Yakunina, G.V. 2014, {\it{J. Astrophys. Astron.}},
{\bf{35}}, 1.

\bibitem{Metcalfe10}
Metcalfe, T.S., Monteiro, M. et al. 2010, {\it{ApJ}}, {\bf{723}},
1583.

\bibitem{Garcia2010}
 Garcia, R.A., Mathur, S. et al. 2010, {\it{Science}}, {\bf{329}},
 1032.

\bibitem{Kollath09}
Kollath, Z. and Olah, K. 2009, {\it{Astron. and Astrophys.}},
{\bf{501}}, 695.

\bibitem{Vaugan1980}
Vaugan, A.H., Preston, G.W. 1980, {\it{PASP}}, {\bf{92}}, 385.

\bibitem{ Noyes84}
Noyes, R.W., Hartman, L., Baliunas, S.L., Duncan, D.K., Vaugan, A.H.
and Parker, E.N. 1984, {\it{ApJ}}, {\bf{279}}, 763.

\bibitem{ Scargle82}
Scargle, J.D. 1982,  {\it{ApJ}}, {\bf{263}}, 835.

\bibitem{Bruevich14b}
 Bruevich, E.A., Bruevich, V.V. and Yakunina, G.V. 2014, {\it{Sun and Geosphere}}, {\bf{8}},
 91.


\bibitem{Olah09}
 Olah, K., Kollath, Z., Granzer, T. et al. 2009, {\it{Astron. and Astrophys.}}, {\bf{501}},
 703.

\bibitem{Morgenthaler11}
Morgenthaler, A., Petit, P., Morin, J., Auriere, M., Dintrans, B.,
Konstantinova-Antova, R., Marsden S. 2011, {\it{Astron. Nachr.}},
{\bf{332}}, 866.



\end{thebibliography}
\end{document}